**Groups of motions and mechanics I: point mechanics**

**D. H. Delphenich** *

It is shown that physical mechanics for pointlike bodies can be effectively modeled in terms of the action of transformation groups that act as symmetries of the solutions of systems of differential equations that describe the integrability of dynamical states.   The equations of motion are then obtained from these integrability equations for the dynamical states.   It is also observed that the functions that define the components of a dynamical state represent a set of mechanical constitutive laws.   The variational formulation of mechanics is shown to be a specialization of these principles.

<div align="center">Contents        Page</div>

* david_delphenich@yahoo.com

## 1 Introduction

One is first led to believe that the soul of mechanics is in systems of differential equations, ordinary in the case of point mechanics and partial in the case of continuum mechanics. At first, the origin of the system is attributed to Newton's second law of motion, but eventually, one is led to believe that Hamilton's least action principle is a more far-reaching way of characterizing the foundations of physical motion. Eventually, one is introduced to the notion that groups – in particular, groups of transformations – play a fundamental role in every branch of physics, including mechanics.

From a purely mathematical perspective, it is possible to combine the two theories of groups of transformations and systems of differential equations by examining the groups of transformations that act on a space in which the solutions of such a system are found and which take solutions to other solutions. Such transformations are generally referred to as *symmetries* of the system of equations. The study of symmetries of differential equations was, in fact, the motivation for Marius Sophus Lie to found what is now incorrectly referred to as the study of Lie groups and Lie algebras. Interestingly, because he was defining everything in terms of local coordinates, by modern standards, what he was defining were what are now referred to as Lie *pseudogroups*. The field of symmetries of differential equations was then expanded in the last century by a long succession of distinguished mathematicians, such as Cartan, Vessiot, and many others [1].

In order to apply the methods of symmetries of systems of differential equations to the foundations of physical mechanics, it is essential to understand that physical motion always involves more than just the association of points in a configuration manifold with other points, as one would derive from a group action on the manifold. Rather, this abstract geometric association of points must be combined with an association of physical quantities that are attached to the points, such as scalars, vectors, tensors, spinors, and the like. These objects are most conveniently modeled as sections of fiber bundles, which are usually vector bundles in the physical applications. Moreover, the sections are generally required to be solutions of some system of partial differential equations, which either gets modeled as an exterior differential system on the total space of the bundle or as a differential operator on the sections. It then becomes clear how motion is related to symmetries of differential equations. Of course, the main question then becomes that of characterizing the nature of the system of differential equations.

The most elementary system that one is introduced to is, of course, Newton's second law of motion, which can be given the form $\ddot{x}^i(\tau) = F^i(\tau, x^i(\tau), \dot{x}^i(\tau))$, which seems to mix kinematical and dynamical states, $f^i(\tau, x^i(\tau), \dot{x}^i(\tau), \ddot{x}^i(\tau)) = 0$, which is more homogeneous as a statement of 2-jets of kinematical states, or $F_i = dp_i/dt$, which involves only dynamical states directly.

Since the form of Newton's equations of motion is generally introduced for translational motion, one then learns how to adapt it to motions that are due to a non-Abelian group in the form of the rotation group, or later, the Lorentz group. On finds that depending upon whether one considers the kinematical and dynamical variables in an inertial frame or a non-inertial one, one might also have to adapt one's definition of the proper time derivative.

---

[1] For a fascinating discussion of the evolution of the field, one can peruse the introductory chapter to Pommaret [1]. For other treatments of the theory of symmetries of differential equations, one can confer Olver [2, 3] or Bluman and Anco [4].

If one starts with a variational basis for mechanics then the first system of differential equations that one encounters is the system of Euler-Lagrange equations that are defined by a choice of Lagrangian for the mechanical system in question. Hence, it is natural to examine the transformations that take solutions of that system to other solutions.

Actually, it is more customary to examine symmetries of the action functional or the Lagrangian density function. Although one is usually introduced to the idea of Noether symmetries in this way, one should keep in mind such symmetries do not exhaust the symmetries of the Euler-Lagrange equations. Nowadays, the role of non-Noether symmetries and conservation laws in physical systems is a fairly well-established concept, especially in the context of nonlinear wave equations.

Furthermore, not every system of differential equations, including many that describe physical systems, can be given a Lagrangian formulation. For instance, dissipative systems, and non-conservative systems, in general, do not admit such a formulation. Hence, one wonders if it possible to find some way of formulating the laws of motion in a manner that is more general than the variational formulation, but not so abstract that it is devoid of physical intuition.

Since the methodology of groups of transformations acting as groups of motions is, perhaps, better established in the context of the symplectic approach to Hamiltonian mechanics (cf., e.g., Souriau [5]. Arnol'd [6], or Abraham and Marsden [7]) than in the context of Lagrangian mechanics, our reason for not starting with the former mathematical methodology must be given. Basically, it comes down to this: Since the Legendre transformation is invertible, a choice of Hamiltonian is essentially equivalent to a choice of Lagrangian, but the methodology that is associated with symmetries of systems of differential equations is more naturally formulated in the language of jet bundles.

It is the purpose of the present study to present the hypothesis that a useful generalization of the least-action principle is to be found in the fact that when one is given a Lagrangian function $\mathcal{L}$ on the bundle $J^k(K, M)$ of jets of whatever objects $K$ in a configuration manifold $M$ that one describing the motion of (e.g., curve segments, compact connected submanifolds of higher dimension than one) its exterior derivative $d\mathcal{L}$ defines a particular type of vertical 1-form on $J^k(K, M)$, namely, an exact one. Since exactness is related to the conservative character of the forces that are associated with the system, and not all forces are conservative, a reasonable generalization would be to vertical 1-forms that are not necessarily exact.

Since this suggests that we are no longer basing our mechanical model in the least action principle, or even a Lagrangian, the question then arises whether one can still define a unique system of equations for a vertical 1-form $\phi$ on $J^k(K, M)$ that defines a dynamical state. It is the basic thesis of this work, expanding on general principles put down by Pommaret in [1], that requiring the integrability of the dynamical state implies a system of differential equations that are defined by the dual of the Spencer operator that acts on the dynamical state. These equations can be shown to generalize the Euler-Lagrange equations that one deduces from starting with a Lagrangian.

Furthermore, whereas the approach of Pommaret was still essentially based in a generalization of the least-action principle, we shall take the position here that when one examines the functional form of the components $(F_i, p_i, \ldots)$ of $\phi$, namely:

$$F_i = F_i(u^a, x^i, x^a{}_i, \ldots), p_i = p_i(u^a, x^i, x^a{}_i, \ldots), \ldots, \qquad (1.1)$$

one sees that what is dealing with is a set of mechanical constitutive laws, such as:

$$F_i = -kx_i, \qquad p_i = m_i \dot{x}^i, \qquad (1.2)$$

in the case of a one-dimensional simple harmonic oscillator. Hence, since any Lagrangian $\mathcal{L}$ will define a 1-form by way of the vertical projection of $d\mathcal{L}$, and indeed, such a set of functional relationships are often implicit in the definition of $\mathcal{L}$, but not every set of constitutive laws define a Lagrangian, we see that we have effectively generalized the scope of Lagrangian mechanics to something that includes constitutive laws that do not define Lagrangians, such as viscous drag forces, which generally take the form $F_i = F_i(u^a, x^i, x^a{}_i)$.

We can further specialize the form of the dynamical equations by assuming that the motion of the object $K$ in $M$ is due to the action of a Lie group $G$ of physical motions on $M$, at least locally. Since an action $G \times U \to M$ prolongs to an action $J^k(K, G) \times J^k(K, U) \to J^k(K, M)$, and in a manner that generalizes the methods of rotational mechanics to more general groups, one finds that if one regards the elements of the bundle $J^k(K, U)$ as being *initial* kinematical states, hence, not functions of time or space, then the dynamical equations that one defines on $J^k(K, M)$ can be pulled back to dynamical equations on $J^k(K, G)$.

Mechanics, in general, subdivides into kinematics, dynamics, and statics. However, in the context of point mechanics statics becomes a specialization of dynamics, whereas in the context of continuum mechanics, one can also regard dynamics as statics in $n+1$ dimensions. This is the approach that was taken by the Cosserat brothers [**8**] in their attempt to formulate continuum mechanics on the basis of groups of motions and the methods of moving frames. Hence, the same mathematics could describe either a surface in equilibrium or a filament in motion, a surface in motion or a solid in equilibrium, etc. However, there is a significant difference between the types of initial or boundary-value problems that one can pose in statics versus dynamics.

At the root of every mechanical model there seem to be five fundamental notions:

*i*) A *configuration space M*,

*ii*) A *Lie group of motions G* that acts on it (perhaps only locally),

*iii*) A *kinematical state space* $\Psi$ that is associated with this action,

*iv*) A *dynamical state space* $\Phi$ that is, in some sense, dual to $\Psi$.

*v*) A *constitutive law* that associates dynamical states with kinematical states.

We shall assume that $M$ and both state spaces are represented by smooth manifolds. The kinematical and dynamical state spaces will not generally have the same dimension, though.

We shall think of the fundamental laws of mechanics as then consisting of the constitutive law and an integrability condition on the dynamical state. One can also represent the dynamical integrability condition as a conservation law, which is the case when it takes the form of either the vanishing of the exterior derivative of an $n-1$-form, or, by Poincaré duality, the vanishing of the divergence of the vector field that is Poincaré dual to it.

Hence, the basic set of laws for a mechanical model will take the form:

$$D^{*}\phi = 0, \qquad\qquad \phi = \phi(\psi), \tag{1.3}$$

in which $\phi$ is the dynamical state, $D^{*}$ is generally a differential operator that describes the integrability of $\phi$ in some manner, and the last equation represents the constitutive law. A fundamental point to be made in what follows is that although integrability follows from variational principles, nevertheless, it can be defined as a basic axiom in a manner that goes beyond the scope of variational methodology.

The present work is divided into two Parts: the first one discusses the case of point mechanics, while the second one discusses the mechanics of extended bodies. Ultimately, in Part II the methods of point motion will be seen to be a reduction of the methods of the motion of extended bodies from a multidimensional world-tube to a one-dimensional world line, which then reduces the partial derivatives to total derivatives.

In sections 2 and 3 of the present Part, we briefly summarize the relevant terminology and results from the theories of groups of transformations and the geometry of jet bundles. In particular, we discuss the representation of systems of differential equations and the calculus of variations in the language of jets. We also discuss the way that a group action on a manifold can give rise to an action of a "prolongation" of the group on the bundle $J^{k}(K, M)$ of $k$-jets of the maps of $K$ into $M$ that define the objects in motion. In section 4, we discuss some of the more common groups of motion for point mechanics.

In each case of motion, viz., pointlike and extended bodies, we shall first discuss the basic problem of the mathematical representation of kinematical states, first, in general, and then in the case where motion is due to the action of a group of motions. We then we discuss the role of integrability as it relates to these states in both forms. The definition that we choose for a dynamical state that is dual to the space of kinematical states is then motivated by examining what the variational formalism would suggest and what would represent a physically interesting generalization of its scope. The integrability of dynamical states is then examined and shown to define a reasonable generalization of the Euler-Lagrange equations that would follow from a choice of Lagrangian. We also discuss the nature of mechanical constitutive laws and show how the general formalism applies to some of the more common physical models for motion.

### 2  Groups of transformations [2]

A Lie group $G$ acts as a group of transformations on a differentiable manifold $M$ if there is a smooth map $G \times M \to M$, $(g, x) \mapsto gx$ that satisfies the semi-group property that $(gg', x)$ goes to $g(g'x)$ for any $g, g' \in G$ and any $x \in M$, and if $e \in G$ is the identity element then $ex = x$ for all $x \in M$.  As a consequence, one has that $g^{-1}(gx) = x$ for any $g \in G$, $x \in M$.

Each $g \in G$ defines a diffeomorphism $L_g$: $M \to M$, $x \mapsto gx$ that one calls *left translation by* $g$; its inverse is then $L_g^{-1} = L_{g^{-1}}$.  One can then define a map $L$: $G \to$ Diff($M$), $g \mapsto L_g$ , where Diff($M$) is the group all diffeomorphism of $M$.  From the semi-group property of any action of G on $M$, it is then a group homomorphism, and its image $L(G)$ is a subgroup of Diff($M$).  Conversely, any subgroup $G$ of Diff($M$) acts on $M$ as a group of transformations in the obvious way: $G \times M \to M$, $(f, x) \mapsto f(x)$.

These last two remarks seem to suggest that the scope of all group actions on a manifold $M$ is identical to the scope of all subgroups of Diff($M$).  However, the group Diff($M$) can be quite nebulous and intractable to start with if one expects to deduce any results that are of a specific nature, as one might hope for in the context of physical mechanics.  For instance, its one-parameter subgroups basically amount to flows on $M$, and $M$ might not admit global flows, but only local ones that are defined for finite time intervals.  Hence, the study of the subgroups of Diff($M$) already includes the study of dynamical systems on $M$, which is quite broad in its own generality.  Furthermore, except in special cases, such as compact $M$, the group Diff($M$) can be regarded as an infinite-dimensional manifold with a group structure, but not an actual Lie group; i.e., the group operations are not differentiable.  (See the discussion of infinite-dimensional Lie groups in Pressley and Segal [**10**].)

Now, the scientific method differs from the mathematical method in various crucial ways, including the fact that the mathematical models for physical phenomena are constructed "from the ground up," not "from the top down," as in mathematics.  That is, to paraphrase Hermann Minkowski: "They are rooted in the soil of experimental physics, and therein lies their strength."  Hence, rather than trying to establish the full scope of all subgroups of Diff($M$), we shall accept the restriction of generality that is implied by considering only specific group actions.

One of the first issues that one must address for a given action of a Lie group $G$ on a manifold $M$ is the extent to which the group $G$ "moves" a given element $x \in M$.  The set $G(x) = \{gx \mid g \in G\}$ is called the *orbit* of $x$ under the action of $G$.  If $G(x) = M$ for some – hence, any – $x \in M$ then the action is called *transitive*.  Otherwise stated, when a group action is transitive, for any pair of points $x, y \in M$ there is at least one $g \in G$ such that $y = gx$.  When there is a transitive action of a Lie group $G$ on a manifold $M$ one calls $M$ a *homogeneous space*.  For instance, by definition, an $n$-dimensional affine space $A^n$ admits a transitive action of the translation group $\mathbb{R}^n$ and any $n$-sphere $S^n$ admits a transitive action of $SO(n+1)$.

---

[2] For a good review of the theory of transformation groups, as applied to physics, one can consult Michel [**9**].

In any event, for a given $x \in M$, one can reduce the representation of $G$ in Diff($M$) to a representation in Diff($G(x)$).

Although we said that there is *at least one* $g \in G$ that takes any $x$ to $y \in G(x)$, we said nothing about its uniqueness. One sees that this ambiguity is related to the number of elements of $G$ that fix the point $x$. In general, the set $G_x = \{g \in G| \ gx = x\}$ is a subgroup of $G$ that one calls the *isotropy subgroup*[3] at $x$ for the action. One immediately finds that all points of an orbit will have isomorphic isotropy subgroups, although the isomorphism is defined by conjugation, and is not unique. In fact, one finds that any orbit can be expressed, as a manifold, as the coset space $G/G_x$, which is also a homogeneous space; i.e., they are diffeomorphic.

In the case of $A^n$, the action of $\mathbb{R}^n$ is *effective*, or *simply transitive*, which means that $G_x = \{e\}$ at every $x \in A^n$, so the translation that takes any $x$ to any $y$ is unique and $A^n$ is diffeomorphic to $\mathbb{R}^n$. In the case of $S^n$, the isotropy subgroup at any point is $SO(n)$ and $S^n$ is then diffeomorphic to $SO(n+1)/SO(n)$. When the action of a group $G$ on a manifold $M$ is effective, the representation of $G$ in Diff($M$) is faithful; i.e., an injective homomorphism.

In the extreme case when $G_x = G$, one calls $x$ a *fixed point* of the action of $G$. There are no fixed points for any effective action, such as the action of the translations on an affine space. If one regards a (global) flow on a manifold $M$ as an action $\Phi: \mathbb{R} \times M \to M$, $(\tau, \ x) \mapsto \Phi_\tau(x)$ then by differentiation one obtains a (velocity) vector field $\mathbf{v}(x) = d\Phi_\tau(x)/d\tau|_{\tau=0}$ and the fixed points of the flow correspond to the zeroes of $\mathbf{v}$. In this example, one also needs to consider the possibility that the isotropy subgroup might be $\mathbb{Z}$, which leads to existence of *periodic orbits* of the flow, which are then diffeomorphic to $\mathbb{R}/\mathbb{Z} = S^1$. More generally, an action of $\mathbb{R}^n$ on $M$ might have an integer lattice $\mathbb{Z}^n$ in $\mathbb{R}^n$ as an isotropy subgroup for some orbits, so the orbits in question would be diffeomorphic to $n$-dimensional torii; this example has an immediate application to the concerns of crystallography.

### 3 Jet bundles [4]

The methods of jet bundles can be applied to two of the most fundamental branches of mathematics, as far as physical models are concerned: differential equations and the calculus of variations. Furthermore, they represent a natural generalization of the concept of a Taylor series expansion of an analytic function on $\mathbb{R}^n$ to $C^k$ functions on an $n$-dimensional differentiable manifold $M$. They also play an important role in the

---

classification of singularities of smooth functions on $M$, although we shall not discuss that topic in this study.

### 3.1  Jets of mappings

Quite simply, the $k$-jet $j^k f|_u$ of a $C^k$ map $f: M \rightarrow N$ from a manifold $M$ to a manifold $N$ at a point $u \in M$ is the equivalence class of all $C^k$ functions that are defined in some neighborhood of $u$ and have the same values for $f(u)$ and their first $k$ derivatives at $u$. Hence, if $U \subset M$ is neighborhood of $u$ on which one has coordinates $u^a$ and $V \subset N$ is a neighborhood of $f(u)$ on which one has defined coordinates $x^i$ then $j^k f|_u$ can be associated with the element of $J^k(\mathbb{R}^m, \mathbb{R}^n) = \mathbb{R}^m \times \mathbb{R}^n \times (\mathbb{R}^{m*} \otimes \mathbb{R}^n) \times (S^2(\mathbb{R}^{m*}) \otimes \mathbb{R}^n) \times \ldots \times (S^k(\mathbb{R}^{m*}) \otimes \mathbb{R}^n)$ that takes the form $(u^a, x^i, x^i_{,a}, x^i_{,a,b}, \ldots, x^i_{,a_1 \cdots, a_k})$ . The notation $S^k(\mathbb{R}^{m*})$ refers to the vector space of completely symmetric covariant tensors over $\mathbb{R}^m$ of degree $k$, so the vector space $S^k(\mathbb{R}^{m*}) \otimes \mathbb{R}^n$ serves as a model space for the space of $k^{\text{th}}$ partial derivatives of $x^i = x^i(u^a)$ with respect to the $u^a$. It is important to see that the coordinates of this element in $J^k(\mathbb{R}^m, \mathbb{R}^n)$ are numbers, not functions, since there will be an infinitude of functions defined in a neighborhood of $u$ that differentiate to the same numbers at $u$.

The set $J^k_u(M, N)$ of all $k$-jets of $C^k$ maps from $M$ to $N$ at $u \in M$ is a manifold that can be associated with $u$, $f(u)$, and $(u, f(u))$. Hence, the disjoint union $J^k(M, N)$ of all $J^k_u(M, N)$ over all $u \in M$ becomes a fibered manifold over $M$ by the projection that takes any $j^k f|_u$ to $u$, a fibered manifold over $N$ by the projection that takes $j^k f|_u$ to $f(u)$ and a fibered manifold over $M \times N$ by the projection that takes $j^k f|_u$ to $(u, f(u))$.

If one looks at the projection $J^k(M, N) \rightarrow J^{k-1}(M, N)$, that locally takes $(u^a, x^i, x^i_{,a}, x^i_{,a,b}, \ldots, x^i_{,a_1 \cdots, a_{k-1}}, x^i_{,a_1 \cdots, a_k})$ to $(u^a, x^i, x^i_{,a}, x^i_{,a,b}, \ldots, x^i_{,a_1 \cdots, a_{k-1}})$, one finds that the fibers of this projection are affine spaces that are modeled on the vector space $S^k(\mathbb{R}^{m*}) \otimes \mathbb{R}^n$, which is essentially the space of all $x^i_{,a_1 \cdots, a_k}$.

A section of the bundle $J^k(M, N) \rightarrow M$ takes the local coordinate form $(u^a, x^i(u), x^i_a(u), \ldots, x^i_{a_1 \cdots a_k}(u))$. Here, the lower indices do not have to represent derivatives.

### 3.3  Jets and power series

By the use of local coordinates, it is straightforward to see how the $k$-jet $f^k|_u$ at $u \in M$ that takes the local form $(u^a, x^i(u), x^i_a(u), \ldots, x^i_{a_1 \cdots a_k}(u))$ is associated with the set of $n$ $k^{\text{th}}$-degree polynomials in $m$ variables:

$$f^i(u^a) = x^i + x^i_a u^a + \ldots + \frac{1}{k! \cdots k!} x^i_{a_1 \cdots a_k} u^{a_1} \cdots u^{a_k} . \tag{3.1}$$

Since the coordinates of $j^k|_u$ are constants, these polynomial functions represent a generalization of the $k^{\text{th}}$-degree Taylor series approximation to a function that is defined in a neighborhood of the origin of $\mathbb{R}^m$.

Now suppose $f: U \to N$ is a $C^k$ function that is defined on some open subset in $M$; hence, it will have a $k$-jet prolongation $j^k f: U \to J^k(U, N)$. If $U$ carries a coordinate system $u^a$ and $f(U)$ is contained in a chart $(V, x^i)$ then $f$ defines a set of $n$ polynomials of degree $k$ in $m$ variables:

$$f^i(u^a) = x^i + x^i_{,a} u^a + \ldots + \frac{1}{k! \cdots k!} x^i_{,a_1,\cdots,a_k} u^{a_1} \cdots u^{a_k} , \tag{3.2}$$

that can be interpreted as the $k^{\text{th}}$-degree Taylor series approximation to $f$ on $U$. However, we see that not every polynomial of the form (3.1) takes the form (3.2), but only the ones that describe *integrable* sections of $J^k(U, N) \to U$. That is, the coefficient functions must satisfy:

$$x^i_{a_1 \cdots a_l} = x^i_{a_1 \cdots a_{l-1}, a_l} , \qquad l = 1, \ldots, k. \tag{3.3}$$

If $f: U \to N$ is a smooth function then it has continuous derivatives of all orders, which means that the power series (3.2) can extend to an infinite series; i.e., one defines a section of $U \to J^\infty(U, N)$ by prolongation. However, although $f$ may be well-defined at every point of $U$, the power series that it generates does not have to converge at every point of $U$, unless $f$ is also analytic. In general, a power series that does not have to converge is referred to as a *formal power series*. Furthermore, not every formal power series whose coefficients are differentiable functions on $U$ necessarily represents a smooth function on $U$, but only the integrable ones; i.e., the recursion (3.3) must extend to infinity.

### 3.2 Integrability of sections of jet bundles

The reason that we did not use commas in the lower indices in expressing the local form for the general section of $J^k(M, N) \to M$ is because not all sections of this bundle take the form of *$k$-jet prolongations* of maps from $M$ to $N$, which then take the local form $(u^a, x^i(u), x^i_{,a}(u), \ldots, x^i_{,a_1,\cdots,a_k}(u))$. That is to say, not all sections of $J^k(M, N) \to M$ are *integrable*. The integrability conditions for a section $f^k: M \to J^k(M, N)$ then take the local form of a set of partial differential equations in the coordinates of $f^k$:

$$x^i_a = x^i_{,a} , \qquad \ldots, \qquad x^i_{a_1 \cdots a_k} = x^i_{a_1 \cdots a_{k-1}, a_k} . \tag{3.4}$$

These equations recursively say that successively higher-degree coordinates must be the partial derivatives of the previous-degree coordinates. Note that this process makes sense only for sections of $J^k(M, N) \to M$, not the individual elements of $J^k(M, N)$, since differentiation is involved.

There are two ways of characterizing integrable sections of $J^k(M, N) \rightarrow M$ depending upon whether one prefers to think in terms of exterior differential systems on $J^k(M, N)$ or differential operators on sections of that bundle over $M$.

In the former case, one first defines the *contact form* on $J^k(M, N)$, which is a 1-form $\Theta$ on $J^k(M, N)$ with values in the vector bundle $V(J^{k-1}(M, N))$ of tangent vectors to $J^{k-1}(M, N)$ that are vertical for the projection onto $M$; that is, they project to zero. Locally, it takes the form of a set of 1-forms for each coordinate of an element of $J^k(M, N)$ past the $u^a$ coordinates and not including the highest-order coordinates:

$$\Theta = (\Theta^i, \Theta^i_a, \ldots, \Theta^i_{a_1 \cdots a_{k-1}}) \tag{3.5}$$

in which:

$$\Theta^i = dx^i - x^i_a du^a, \tag{3.6a}$$

$$\Theta^i_a = dx^i_a - x^i_{ab} du^b, \tag{3.6b}$$

$$\ldots,$$

$$\Theta^i_{a_1 \cdots a_{k-1}} = dx^i_{a_1 \cdots a_{k-1}} - x^i_{a_1 \cdots a_k} du^{a_k}). \tag{3.6c}$$

The section $\Theta$ of $\Lambda^1(J^{k-1}(M, N)) \otimes V(J^{k-1}(M, N))$ is then locally represented by:

$$\Theta = \Theta^i \otimes \frac{\partial}{\partial x^i} + \Theta^i_a \otimes \frac{\partial}{\partial x^i_a} + \cdots + \Theta^i_{a_1 \cdots a_k} \otimes \frac{\partial}{\partial x^i_{a_1 \cdots a_k}}. \tag{3.7}$$

One immediately notes that for a $k$-jet prolongation of a map $x: M \rightarrow N$, for which the coordinates of the form $x^i_{a_1 \cdots a_s}$ are partial derivatives of $x^i$, all of the 1-forms in (3.5) vanish. Indeed, the converse is also true. Hence, a section $f^k: M \rightarrow J^k(M, N)$ is integrable iff the 1-form $(f^k)^* \Theta$ on $M$ that is obtained by pulling back $\Theta$ by the section $f^k$ vanishes.

One can identify an important class of transformations of the manifold $J^k(M, N)$ in the form of the *contact transformations*, which preserve the 1-form $\Theta$. That is, if $\Phi: J^k(M, N) \rightarrow J^k(M, N)$ is a fiber-preserving diffeomorphism then $\Phi$ is a contact transformation iff $\Phi^* \Theta = \Theta$. Such a transformation will then take integrable sections of $J^k(M, N) \rightarrow M$ to other integrable sections.

By differentiation, one can also define the infinitesimal generators of one-parameter families of contact transformations. An *infinitesimal contact transformation* will then be a vector field $\mathbf{X}$ on $J^k(M, N)$ such that:

$$0 = L_{\mathbf{X}} \Theta = d i_{\mathbf{X}} \Theta + i_{\mathbf{X}} d\Theta. \tag{3.8}$$

If one prefers to deal with differential operators on sections of $J^k(M, N) \rightarrow M$ then an equivalent way of characterizing integrable sections is by means of the *Spencer operator*, which takes the form $D: J^k(M, N) \rightarrow T^*(M) \otimes J^{k-1}(M, N), f^k \mapsto j^1(f^{k-1}) - f^k$. It was first defined by Donald Spencer in his work on the deformations of structures defined by pseudogroups, and then applied to the formal integrability of systems of linear partial differential equations [**13**]. (For the nonlinear case, see Goldschmidt [**14**].)

Perhaps the best way to interpret the Spencer operator locally on a coordinate chart $(U, u^a)$ is to regard an element of $T^*(M) \otimes J^{k-1}(M, N)$ as a 1-form $\omega = \omega_a(u)\, du^a$ on $U$ whose components $\omega_a(u)$ are expressed as the $(k-1)^{\text{th}}$-degree Taylor series in the variables $u^a$ that is defined by a section $f^{k-1}$ of the fibration $J^{k-1}(M, N) \to M$. Hence, it is unambiguous what we would mean by the notation $\left( u^a, y^i(u^a), \cdots, y^i_{a_1 \cdots a_{k-1}}(u^a) \right) du^a$ for a local section of $T^*(M) \otimes J^{k-1}(M, N) \to U$. When a section of $J^k(M, N) \to U$ has the local form $f^k(u^a) = (u^a, x^i(u), x^i_a(u), \ldots, x^i_{a_1 \cdots a_k}(u))$, the operator $D$ then gives:

$$Df^k = \left( u^a, Dx^i, \cdots, Dx^i_{a_1 \cdots a_{k-1}} \right) du^a \ . \tag{3.9}$$

in which:

$$Dx^i_{a_1 \cdots a_m} = \frac{\partial x^i_{a_1 \cdots a_{m-1}}}{\partial u^{a_m}} - x^i_{a_1 \cdots a_m} \ . \tag{3.10}$$

The relationship between this operator and the contact form is given by:

$$\Theta^i_{a_1 \cdots a_m} = Dx^i_{a_1 \cdots a_m}\, du^{a_m} \ . \tag{3.11}$$

Clearly, $f^k$ is integrable iff:

$$Df^k = 0 \ . \tag{3.12}$$

A contact transformation $\Phi\colon \Gamma(M, J^k(M, N)) \to \Gamma(M, J^k(M, N))$, which we now understand to mean an invertible map on sections of the bundle $J^k(M, N) \to M$, can then be characterized by the property that a section $\psi \in \Gamma(M, J^k(M, N))$ is integrable iff $\Phi(\psi)$ is integrable; thus, $D\psi = 0$ iff $D(\Phi(\psi)) = 0$.

The $D$ operator can be extended to an operator $D\colon \Lambda^l(M) \otimes J^{k-l}(M, N) \to \Lambda^{l+1}(M) \otimes J^{k-l-1}(M, N)$ by setting:

$$D(\omega \otimes f^{k-l}) = d\omega \otimes f^{k-l-1} + \omega \otimes Df^{k-l} \ . \tag{3.13}$$

One then sees that $D^2 = 0$, and the resulting sequence:

$$C^k(M, N) \xrightarrow{\ j_k\ } J^k(M, N) \xrightarrow{\ D\ } T^*(M) \otimes J^{k-1}(M, N) \xrightarrow{\ D\ } \Lambda^2(M) \otimes J^{k-2}(M, N) \xrightarrow{\ D\ },$$

which terminates when $l$ reaches either $m = \dim(M)$ or $k$, is exact; i.e., the image of any map is the kernel of the one that follows it [5].

---

[5] We are implicitly treating all terms in the sequence past the first as spaces of sections, and all bundles past the second one as vector bundles.

### 3.4 Systems of differential equations

When jets are expressed in local form it is straightforward to define a system of differential equations, whether ordinary or partial, in terms of jets of mappings. Namely, if $f: M \to N$ is a $C^k$ mapping and its $k$-jet prolongation $j^k f: M \to J^k(M, N)$ has the local form $j^k f = (u^a, x^i(u), x^i_{,a}(u), \ldots, x^i_{,a_1 \cdots a_k}(u))$ then if $F: J^k(M, N) \to \mathbb{R}$ is any function, one can define a homogeneous differential equation of order $k$ by way of:

$$0 = F(j^k f) = F(u^a, x^i, x^i_a, \ldots, x^i_{a_1 \cdots a_k}). \qquad (3.14)$$

Now, $F$ is differentiable and $dF \neq 0$ then $F^{-1}(0)$ is a submanifold of $J^k(M, N)$. Although the submanifold $F^{-1}(0)$ projects onto $M$, $N$, and $M \times N$, it is not necessarily a fiber bundle over these spaces, but only a *fibered submanifold*, which means that the projection map is a surjective submersion. When $J^k(M, N) \to M$ is a vector bundle, one can speak of the linearity of $F$, and if $F$ is linear by restriction to the fibers then $F^{-1}(0)$ will consist of vector spaces fibered over $M$, but not necessarily a vector sub-bundle of $J^k(M, N)$. In such an event, $F$ defines a linear differential equation of order $k$.

One can characterize a *solution* of the differential equation (3.14) as a $C^k$ map $f: M \to N$ such that $j^k f$ satisfies $F(j^k f) = 0$. Since $j^k f$ is an integrable section of $J^k(M, N) \to M$, one sees that any fiber-preserving diffeomorphism $\Phi: J^k(M, N) \to J^k(M, N)$ that takes solutions of (3.14) to other solutions must be, above all, a contact transformation. Furthermore, it must preserve $F$, in the sense that $F \cdot \Phi = F$. One calls such transformations *symmetries* of the differential equation defined by $F$.

If one expands $dF$ with respect to a local coordinate system then one gets:

$$dF = F_{,a} du^a + F_{,i} dx^i + F_i^a dx^i_a + \cdots + F_i^{a_1 \cdots a_k} dx^i_{a_1 \cdots a_k}, \qquad F_i^{a_1 \cdots a_s} \equiv \frac{\partial F}{\partial x^i_{a_1 \cdots a_s}}. \qquad (3.15)$$

The coefficient of the last term has a special significance and is referred to as the *symbol* of $F$. One can also characterize it in a manner that is independent of the choice of coordinate system by saying that it is the restriction of $dF$ to the vertical sub-bundle of $T(J^k(M, N))$ under the projection of $J^k(M, N)$ onto $J^{k-1}(M, N)$, which takes $(u^a, x^i, x^i_a, \ldots, x^i_{a_1 \cdots a_k})$ to $(u^a, x^i, x^i_a, \ldots, x^i_{a_1 \cdots a_{k-1}})$, locally.

One can generalize (3.14) in various ways: For instance, one can choose other real numbers besides zero, and as long as $dF$ is non-vanishing for those values one can define inhomogeneous differential equations of order $k$. Similarly, one can define $r$ functions $F^r$, $r = 1, \ldots, \rho$ and obtain a system of $r$ differential equations of order $k$, or equivalently, a function $F: J^k(M, N) \to \mathbb{R}^r$.

Since one usually expects components in $\mathbb{R}^r$ to come from elements of $r$-dimensional real vector spaces or manifolds, by way of frames or coordinates, respectively, one can generalize $\mathbb{R}^r$ to a manifold $V$, or, more generally, the fibers of a bundle $B \to M$. This allows one to represent a differential operator $\mathcal{D}: \Gamma(E) \to \Gamma(B)$ that takes sections of one bundle $E \to M$ to sections of another bundle $B \to M$ as being an operator that factors through $\mathcal{D}\phi = F \cdot j_\phi^k f$, for some bundle map $F: J^k(M, N) \to B$ when $\phi \in \Gamma(E)$.

### 3.5  Calculus of variations [6]

In addition to defining a differential equation of order $k$, a differentiable function $\mathcal{L}$: $J^k(M, N) \to \mathbb{R}$ defines a *Lagrangian density* for a class of variational problems. Namely, if $K \subset \mathbb{R}^m \to M$ is a compact orientable $m$-dimensional submanifold with boundary (more generally, a differentiable singular $m$-chain) in an $n$-dimensional manifold $M$ and $V \in \Lambda^m(\mathbb{R}^m)$ is a volume element on $\mathbb{R}^m$ then one can pull $V$ up to an $m$-form on $J^k(K, M)$ by way of the projection on $K$. One can then define an *action functional* on the $C^k$ maps $f: K \to M$ by way of:

$$S[f] = \int_K \mathcal{L}(j^k f)V = \int_K \mathcal{L}(u^a, x^i(u), x^i_{,a}(u), \cdots x^i_{,a_1,\cdots,a_k}(u))V \;. \tag{3.16}$$

If $\delta\!\!\!/f$ is a vector field on $f(K)$, which we think of as an infinitesimal generator of a differentiable homotopy of $f$ and refer to as a *variation* of $f$, then we define the induced variation of $S$ by:

$$\delta\!\!\!/S[\delta\!\!\!/f] = \int_K L_{\delta^k f}(\mathcal{L}(j^k f)V) = \int_K (i_{\delta^k f} d\mathcal{L})V \;. \tag{3.17}$$

In this expression, $L$ refers to the Lie derivative of the $m$-form $\mathcal{L}(j^k f)V$ with respect to the vector field $\delta\!\!\!/^k f$ on $J^k(K, M)$, which is the $k^{\text{th}}$ prolongation of $\delta\!\!\!/f$ and takes the local form:

$$\delta\!\!\!/^k f = \delta x^i \frac{\partial}{\partial x^i} + \frac{\partial(\delta x^i)}{\partial u^a}\frac{\partial}{\partial x^i_a} + \cdots \frac{\partial(\delta x^i)}{\partial u^{a_1}\cdots\partial u^{a_k}}\frac{\partial}{\partial x^i_{a_1\cdots a_k}} \;. \tag{3.18}$$

This makes:

$$i_{\delta^k f} d\mathcal{L} = \delta x^i \frac{\partial\mathcal{L}}{\partial x^i} + \frac{\partial(\delta x^i)}{\partial u^a}\frac{\partial\mathcal{L}}{\partial x^i_a} + \cdots \frac{\partial(\delta x^i)}{\partial u^{a_1}\cdots\partial u^{a_k}}\frac{\partial\mathcal{L}}{\partial x^i_{a_1\cdots a_k}} \;. \tag{3.19}$$

By the usual integration by parts argument, the variation of $S$ by $\delta\!\!\!/f$ takes on the form:

---

$$\delta S[\delta f] = \int_K \frac{\delta L}{\delta f}(\delta f)V + \int_{\partial K} \Theta(\delta f) \,, \tag{3.20}$$

in which $\delta \mathcal{L}/\delta f \in \Lambda^1(f(M))$ is the *variational derivative* of $\mathcal{L}$ with respect to $f$ and $\Theta \in \Lambda^1(f(M)) \otimes \Lambda^{m-1}(K)$. To first order ($k = 1$), one has:

$$\frac{\delta \mathcal{L}}{\delta f} = \left(\frac{\partial \mathcal{L}}{\partial x^i} - \frac{\partial}{\partial u^a}\frac{\partial \mathcal{L}}{\partial x_a^i}\right)dx^i \,, \quad \Theta = \frac{\partial \mathcal{L}}{\partial x_a^i}dx^i \otimes i_{\partial_a}V \,. \tag{3.21}$$

The classical variational problems that are defined by $\mathcal{L}$ and $K$ are the fixed-boundary and variable-boundary problems. In either case, one looks for the *extremal* maps $f \colon K \to N$, namely, the ones that have the property that $\delta S[\delta f] = 0$ for any $\delta f$ of the specified type. For a fixed-boundary problem, that type is defined by variations of $f$ that vanish on the boundary of $f(K)$. For a variable-boundary problem, one weakens this to all variations that satisfy the *transversality condition* that $\Theta(\delta f) = 0$ on the boundary of $f(K)$.

In either type of problem, an extremal must satisfy the *Euler-Lagrange* equations:

$$\frac{\delta \mathcal{L}}{\delta f} = 0 \,. \tag{3.22}$$

### 3.6 Prolongations of group actions

Suppose that $U \subset M$ and one has a group action $G \times U \to M$, $(g, x) \mapsto gx$, and $C^k$ maps $g \colon K \to G$, $u \mapsto g(u)$ and $x_0 \colon K \to U$, $u \mapsto x_0(u)$. By repeated differentiation, one can obtain an action $J^k(K, G) \times J^k(K, U) \to J^k(K, M)$, $(j^k g, j^k x_0) \mapsto j^k x$. In the first two orders, one has:

$$x = gx_0, \tag{3.23a}$$
$$dx = d(gx_0) = dg\,x_0 + g\,dx_0, \tag{3.23b}$$
$$d^2x = d^2(gx_0) = d^2g\,x_0 + 2\,dg\,dx_0 + g\,d^2x_0, \tag{3.23c}$$

(the product $dg\,dx_0$ implicitly means the symmetrized tensor product) so we can define the action of 0-jet sections on 0-jet sections, 1-jet sections on 1-jet sections, and 2-jet sections on 2-jet sections by:

$$(u, g(u)) \times (u, x_0(u)) \mapsto (u, x(u)), \tag{3.24a}$$
$$(u, g(u), dg(u)) \times (u, x_0(u), dx_0(u)) \mapsto (u, x(u), dx(u)), \tag{3.24b}$$
$$(u, g(u), dg(u), d^2g(u)) \times (u, x_0(u), dx_0(u), d^2x_0(u))$$
$$\mapsto (u, x(u), dx(u), d^2x(u)), \tag{3.24c}$$

with the appropriate substitutions from (3.23a, b, c)

So far, we have only described the prolongation of the group action as it affects integrable sections. One sees that (3.23a, b, c) generalize directly to define an action of general elements of $J^2(K, G)$ on elements of $J^2(K, U)$ by way of:

$$x = gx_0, \tag{3.25a}$$

$$\overset{(1)}{x} = \overset{(1)}{g}\, x_0 = \overset{(1)}{g}\, x_0 + g\, \overset{(1)}{x_0}, \tag{3.25b}$$

$$\overset{(2)}{x} = \overset{(2)}{g}\, x_0 = \overset{(2)}{g}\, x_0 + 2\, \overset{(1)}{g}\, \overset{(1)}{x_0} + g\, \overset{(2)}{x_0}, \tag{3.25c}$$

in which we have represented the general elements of $J^2(K, G)$ and $J^2(K, U)$ by $(u, g, \overset{(1)}{g}, \dots, \overset{(k)}{g})$ and $(u, x_0, \overset{(1)}{x_0}, \dots, \overset{(k)}{x_0})$, respectively. We can the general to an action $J^k(K, G) \times J^k(K, U) \to J^k(K, M)$, $(\psi_G, \psi_U) \mapsto \psi$.

Although this form of the action is the most straightforward to explain, nevertheless, when dealing with vectors that are tangent to $G$, it is more mathematically illuminating to left-translate them to elements of the Lie algebra $\mathfrak{g} = T_e G$. Similarly, at the next level of differentiation, one translates the resulting element to $T_0\mathfrak{g}$, and so on. This modifies (3.25a, b, c) to take the form:

$$x = gx_0, \tag{3.26a}$$

$$\overset{(1)}{x} = g(\omega x_0 + \overset{(1)}{x_0}), \tag{3.26b}$$

$$\overset{(2)}{x} = g(\overset{(2)}{\omega} x_0 + 2\omega\, \overset{(1)}{x_0} + \overset{(2)}{x_0}), \tag{3.26c}$$

into which we have introduced:

$$\omega = g^{-1}dg, \tag{3.27a}$$

$$\overset{(1)}{\omega} = \omega\omega + d\omega. \tag{3.27b}$$

As we shall see later, these expressions are at the root of the introduction of angular velocity and angular acceleration, along with the associated Coriolis velocity and accelerations, as well as the normal and centripetal accelerations. However, they also generalize the process beyond the scope of time derivatives of time-varying rotations to partial derivatives of more general group elements acting non-uniformly on initial kinematical states. We shall return to this in Part II.

When the elements of $J^k(K, G)$ have the local form $(u, g, \omega, \overset{(1)}{\omega}, \dots)$ we can modify the prolonged action $J^k(K, G) \times J^k(K, U) \to J^k(K, M)$ as it was described in (3.24a, b, c) to look like:

$$(u, g) \times (u, x_0) \mapsto (u, x), \tag{3.28a}$$

$$(u, g, \overset{(1)}{\omega}) \times (u, x_0, \overset{(1)}{x_0}) \mapsto (u, x, \overset{(1)}{x}), \tag{3.28b}$$

$$(u, g, \overset{(1)}{\omega}, \overset{(1)}{\omega}) \times (u, x_0, \overset{(1)}{x_0}, \overset{(2)}{x_0}) \;\mapsto\; (u, x, \overset{(1)}{x}, \overset{(2)}{x}), \tag{3.28c}$$

for $k = 0, 1, 2$, resp., with the substitutions described in (3.26a, b, c).

What we are really doing locally with the coordinates $\omega, \overset{(1)}{\omega}, \ldots$ in the jet $\psi_G$ is prolonging the Lie algebra $\mathfrak{g}$. The way that one does such a thing in general, at least when $\mathfrak{g}$ acts linearly on a vector space $V$, is as follows:

The first prolongation of $\mathfrak{g}$, which is denoted by $\mathfrak{g}^{(1)}$, consists of those $T \in \mathrm{Hom}(V, \mathfrak{g})$ – i.e., linear maps from $V$ to $\mathfrak{g}$ – such that $T(\mathbf{v})\mathbf{w} = T(\mathbf{w})\mathbf{v}$ for all $\mathbf{v}, \mathbf{w} \in V$. Hence, one can also regard $T$ as an element of $V \otimes \mathfrak{g}$ whose components $T^i_{jk}$ with respect to any frame on $V$ are symmetric in the lower indices; naturally, we are assuming that an element of $a \in \mathfrak{g}$ is represented by a matrix of the form $a^i_j$ relative to such a frame on $V$.

One defines further prolongations of $\mathfrak{g}$ recursively by $\mathfrak{g}^{(k)} = [\mathfrak{g}^{(k-1)}]^{(1)}$. For example, the second prolongation $\mathfrak{g}^{(2)}$ consists of those $T \in \mathrm{Hom}(V, \mathfrak{g}^{(1)})$ such that $T(\mathbf{v})\mathbf{w} = T(\mathbf{w})\mathbf{v}$ for all $\mathbf{v}, \mathbf{w} \in V$. Such a $T$ can be regarded as an element of $S^2(V) \otimes \mathfrak{g}$, viz., a symmetric, second-rank, covariant tensor on $V$ with values in $\mathfrak{g}$.

In general, an element of $\mathfrak{g}^{(k)}$ takes the form of an element of $S^k(V) \otimes \mathfrak{g}$. When one is concerned with $C^k$ maps $g : K \subset \mathbb{R}^m \to G$, the successive derivatives in $j^k g$ take the form of elements in $S^l(\mathbb{R}^m) \otimes \mathfrak{g}$ for $l > 0$. Hence, one can think of the spaces $\mathfrak{g}^{(l)}$, $l = 1, \ldots, k$ in such a case as representing the spaces in which the successive derivatives of $g$ take their values, after left-translation by $g$. One can also form the direct sum $\mathfrak{g}[k] = \mathfrak{g} \oplus \mathfrak{g}^{(1)} \oplus \ldots \oplus \mathfrak{g}^{(k)}$ and obtain the *formal algebra* associated with the $k^{\text{th}}$ prolongation of $\mathfrak{g}$. It is the model vector space for the fibers of the fibration $J^k(K, G) \to K \times G$.

Any important issue to address in the context of prolongations of Lie algebras is whether the process of prolongation goes on to indefinitely high values of $k$ or terminates after a finite number of steps. That is, does there exist some minimum $k$ for which $\mathfrak{g}^{(k+1)}$, and therefore all higher prolongations, vanishes. If such a $k$ exists then $\mathfrak{g}$ is said to be of *finite type* and $k$ is the *type* of $\mathfrak{g}$; otherwise, $\mathfrak{g}$ is of *infinite type*. For example, the first prolongation of $\mathfrak{so}(n)$ vanishes and the second prolongation of the conformal Lie algebra $\mathfrak{co}(n)$ vanishes, so they are of type 1 and 2, respectively. By contrast, $\mathfrak{gl}(n)$ and $\mathfrak{sl}(n)$ are of infinite type.

Since any finite-dimensional Lie algebra is the Lie algebra of some Lie group, the question then arises how one can associate the prolonged Lie algebras with corresponding Lie groups. One can, in fact, prolong the Lie group $G$ that is associated with $\mathfrak{g}$ to begin with in a manner that is consistent with the prolongation of $\mathfrak{g}$. The process basically involves truncated polynomial multiplication, although we shall not elaborate on the details here, but refer the interested reader to Reinhart [**17**]. In the sequel, we shall simply refer to the $k^{\text{th}}$ prolongation of $G$ by $G^{(k)}$ and understand that it is a Lie group that is associated with the Lie algebra $\mathfrak{g}^{(k)}$.

### 3.7 Relationship between jets and other formalisms

There are other ways of characterizing the basic objects of physical mechanics than the methods of jet bundles. The two that we shall mention here are the method of moving frames and the method of Lie groupoids, which are both closely related to the enveloping generality of jets.

The stated purpose of the treatise of the Cosserat brothers [8] on the mechanics of deformable bodies was to apply the method of moving frames that Darboux had used to great advantage in his treatise in the geometry of surfaces. As a result, nowadays one sometimes refers to the bundle of orthonormal frames on a Riemannian manifold as a *Cosserat continuum* or a *Cosserat medium* [7]. Since the method of moving frames was also advocated by Cartan as the basis for differential geometry, this method also has the advantage that it leads into a vast body of literature concerning the Cartan approach to differential geometry.

Although one can define a frame $\mathbf{e}_x$ at a point $x \in M$ as simply a basis for the tangent space $T_x(M)$, for the purpose of relating frames to jets, it is more convenient to represent a frame in $T_x(M)$ as a linear isomorphism $\mathbf{e}_x \colon \mathbb{R}^m \to T_x(M)$. If we now consider the 1-jet $j^1 f \mid_0$ of a local diffeomorphism $f \colon \mathbb{R}^m \to M$ at 0 then we see that it takes the form of precisely such a linear isomorphism. Hence, following Reinhart [17], we can also define the manifold $GL(M)$ of linear frames on $M$ to be the fiber of the bundle $J^1(\mathbb{R}^m, M) \to \mathbb{R}^m$ over 0. The manifold $GL(M)$ is then fibered over $M$. One can then associate the local coordinates $(x^i, g^i_j)$ of a frame in $GL(M)$ with the local coordinates $(u^j, x^i, x^i_j)$ of a 1-jet in $J^1(\mathbb{R}^m, M)$ by setting $u^j = 0$ and $x^i_j = g^i_j$. It is important to note that since the vectors of a frame are linearly independent the corresponding 1-jet must have an invertible coordinate matrix for $x^i_j$, which is why one must restrict to jets of local diffeomorphisms.

One can prolong the definition of a frame on $M$ by defining a frame on $GL(M)$. Hence, such a prolonged frame will represent $m + m^2$ linearly independent vectors tangent to some element $\mathbf{e}_x \in GL(M)$. We then define a bundle $GL^{(1)}(M) \to M$ that we call the *first prolongation* of $GL(M)$.

An important distinction between $GL(M)$ and $GL^{(1)}(M)$ is the fact that, whereas the manifold $M$ does not have to be parallelizable, the manifold $GL(M)$ does. That is, the existence of a linear connection on $GL(M)$ will imply the existence of a global frame field on $GL(M)$, which is usually defined by the $m$ basic horizontal vector fields and the $m^2$ fundamental vertical vector fields (see Kobayashi and Nomizu [20] or Bishop and Crittenden [21]). Hence, although one can think of $GL(M)$ as only locally diffeomorphic to $M \times GL(m)$, one can think of $T(GL(M))$ as globally diffeomorphic to $T(M) \times T(G)$, which also makes $GL^{(1)}(M)$ diffeomorphic to $GL(M) \times GL(m) \times \mathfrak{gl}(m)$. If we denote the first prolongation of the group $GL(m)$ by $GL^{(1)}(m)$ then $GL^{(1)}(m)$ acts on $GL^{(1)}(M)$ just as $GL(m)$ acts on $GL(M)$; indeed, $GL^{(1)}(M)$ is a $GL^{(1)}(m)$-principal bundle.

---

[7] See also the treatment of Cosserat media that is given in Teodorescu [18] or the IUTAM conference proceedings [19]., as well as the discussion in Pommaret [1].

One can iterate this process of prolongation recursively to define the $k^{\text{th}}$ prolongation of $GL(M)$ to be the bundle $GL^{(k)}(M)$ that one obtains from the first prolongation of $GL^{(k-1)}(M)$. Hence, it consists of linear frames in the tangent spaces to the manifold $GL^{(k-1)}(M)$ and defines a $GL^{(k)}(m)$-principal bundle over $M$. It is also diffeomorphic to $GL(M) \times GL(m) \times \mathfrak{gl}^{(k-1)}(m)$.

When one represents 1-frames on $M$ by jets of local diffeomorphisms of $\mathbb{R}^m$ into $M$ at 0, this process of prolongation admits an immediate analogue in the prolongation of jets and jet bundles. One simply represents the manifold $GL^{(k)}(M)$ by the fiber over 0 of $J^k(\mathbb{R}^m, M) \to \mathbb{R}^m$. Once again, it is essential to restrict oneself to jets of diffeomorphisms, which will then have invertible coordinate matrices in first order.

More generally, one usually considers reductions of $GL(M) \to M$ that are defined by choosing some subgroup $G \to GL(m)$. Such reductions are called *G-structures* on $M$ and include such geometrically important cases as the bundle of unit-volume frames defined by a unit-volume element, the bundle of orthonormal frames defined by a metric, the bundle of adapted frames defined by a choice of sub-bundle in $T(M)$, and essentially all of the other geometrically important frame bundles. By the action of the prolongations of $G$ on each $GL^{(k)}(M)$, $k = 1, 2, \ldots$, one defines the prolongations of *G*-structures.

Since we just pointed out that $GL^{(k)}(m)$ acts on $GL^{(k)}(M)$, if a group of motions $G$ is a subgroup of $GL(m)$ then we can consider the motion of frames on $M$ as resulting from the action of $J^k(K, G^{(k)})$ on $GL^{(k)}(U)$ rather than the action of $J^k(K, G^{(k)})$ on $J^k(K, U)$, that describes motion of points of $U \subset M$. Indeed, since any frame on $M$ projects to a point of $M$ it is clear that the motion of frames has more detail to it than the motion of points, due to the fiber dimensions of $GL(M)$. However, our reason for choosing to stay with the methods of more general jets than frames and their prolongations in the present study is that it makes it simpler to discuss the question of integrability.

Although we saw that it is straightforward to represent linear frames and their prolongations by way of jets of local diffeomorphisms of $\mathbb{R}^m$ into $M$, we also see that a *G*-structure on $M$ and its prolongations cannot generally be represented by a bundle of $k$-jets directly. Rather, one must specify a fibered submanifold of $J^k(\mathbb{R}^m, M)$ that is characterized by the solutions to some set of equations on the vertical part of the jets relative to the projection $J^k(\mathbb{R}^m, M) \to (\mathbb{R}^m, M)$, $j^k_x f_x \mapsto (x, f_x)$. For a $k$-jet that is the prolongation of a local diffeomorphism $x^i(u^a)$, the vertical part will be locally characterized by the successive derivatives $(x^i_{,a}, x^i_{,a_1,a_2}, \ldots, x^i_{,a_1 \cdots, a_k})$, so one can think of the fibered submanifold in question as composed of the solutions to a system of $k^{\text{th}}$ order partial differential equations in the functions $x^i(u^a)$

   This brings us to the method of Lie groupoids [8] that has been emphasized by Pommaret [**1**] as the natural setting for the Cosserat approach to mechanics. A Lie groupoid $\Gamma$ differs from a Lie group in several ways: Among them are the fact that composition is not always defined, but behaves like the composition of successive maps, there are two projections $\alpha$: $\Gamma \rightarrow M$ and $\beta$: $\Gamma \rightarrow N$, called the *source* and *target* projections, and there is not a unique identity element, but a left identity associated with every point of $M$ and a right identity associated with every point of $N$. One recovers the notion of a Lie group by considering the *isotropy group* associated with any $(x, y) \in M \times N$, which consists of the fiber over that ordered pair under the projection $(\alpha, \beta)$: $\Gamma \rightarrow M \times N$.

   Perhaps the simplest example of a Lie groupoid, and the local model for all of the others, is $\mathbb{R}^n \times G \times \mathbb{R}^n$, for which the isotropy group at any $(x, y) \in \mathbb{R}^n \times \mathbb{R}^n$ is $G$. The composition of $(x, g, y)$ and $(w, g', v)$ is defined iff $y = w$ and equals $(x, gg', v)$. The left identity at $x$, and the right identity at $y$ are then $(x, e, y)$, in which $y$ or $x$ ranges over all $\mathbb{R}^n$, respectively.

   Another example that is of interest to mechanics, and the one that Pommaret concentrates on, is the Lie groupoid $J^k(M)$ of all invertible $k$-jets of local diffeomorphisms $f$: $U \rightarrow M$. By this, we mean that one composes $k$-jets by the rule $j^k f \cdot j^k f' = j^k(f \cdot f')$, so a $k$-jet is invertible iff there is a $k$-jet $(j^k f)^{-1}$ such that $j^k f \cdot (j^k f)^{-1} = j^k I$. A fibered submanifold of this Lie groupoid, such as one associates with a special class of local diffeomorphisms, then defines a system of $k^{\text{th}}$ order partial differential equations that one refers to as a *Lie equation*. For instance, one can obtain the $k^{\text{th}}$ order Lie equations for local volume-preserving diffeomorphisms by prolonging the basic equation:

$$\det\left(\frac{\partial y^i}{\partial x^j}\right) = 1, \tag{3.29}$$

and the Lie equation for local isometries is obtained by prolonging the basic equation:

$$g_{mn}(x)\frac{\partial y^m}{\partial x^i}\frac{\partial y^n}{\partial x^j} = g_{ij}(x). \tag{3.30}$$

   An example of a Lie groupoid that shows how they relate to $G$-structures is the Lie groupoid of all local $G$-isomorphisms of a $G$-structure $G(M)$. Such an isomorphism is a local diffeomorphism that takes an open subset $U \subset G(M)$ to another open subset of $G(M)$ in a manner that takes elements of one fiber to elements of the same fiber and commutes with the action of $G$. By prolongation, one can define the Lie groupoid of local $G^{(k)}$-isomorphisms of $G^{(k)}(M)$.

   Although there are advantages to the generality that is associated with the methods of Lie groupoids, since our main objective in what follows to focus on the same issues at a more elementary level from the standpoint of mechanics, one must regard the present effort as a reduction in scope from a purely mathematical standpoint. However, since the mathematical methods of Lie groupoids and Lie equations are mostly defined in the context of problems that many pure mathematicians find less than intuitive, we hope that by discussing the foundations of mechanics at a more specialized level perhaps some intuitive appeal can be restored to the more general methods.

---

[8] Besides the book by Pommaret, other good references on Lie groupoids are the thesis of Ngo Van Que [**22**] and the book by MacKenzie [**23**].

## 4 Groups of physical motions

In this part of our study of the formulation of mechanics in terms of groups of motions, we shall briefly summarize the finite-dimensional Lie groups that pertain to the motion of rigid bodies. In the next part of this series, we shall discuss a useful way of extending to an infinite-dimensional group that describes motions of an extended deformable body. Most of the basic mathematical concepts and results in this section can be found in Chevalley [**24**].

The concept of a rigid body is one way of approximating the motion of an extended object along a congruence of curves to the motion of a point along a single curve, or rather, the motion of an orthonormal frame along a curve. By assuming that the object is rigid – i.e., all distances between pairs of points of the object remain constant in time – one replaces the mass density function over the object with a constant – the total mass – at the center of mass and the orthonormal frame field over the object that describes the angular positions of infinitesimal neighborhoods of each point with a single orthonormal frame at the center of mass. Similarly, the translational velocity vector field of the object reduces to the velocity of the curve followed by the center of mass and the angular velocity 1-form for the object reduces to a 1-form on the tangent spaces to that curve.

There are only certain motions of a rigid body that will preserve its rigidity. Since rigidity is a metric concept, if the body moves in a general Riemannian or Lorentzian manifold $(M, g)$ then the rigid motions will be isometries of the metric $g$.

At the most elementary level of non-relativistic mechanics, $M$ is $(\mathbb{R}^n, \delta)$, where $\delta = \delta_{ij}$ $dx^i dx^j$ is the Euclidian metric. It can be shown (cf., Arnol'd [**6**]) that the group of physically meaningful transformations that preserve this metric consists of the semi-direct product $\mathbb{R}^n \lhd SO(n)$ of the translation group with the orientation-preserving rotation group. For $n = 2$, the rotations in question are fixed-axis rotations, so the rotation group $SO(2)$ is Abelian and one-dimensional, but for $n = 3$, the rotational axis can point to anywhere on the unit 2-sphere, so the rotation group is three-dimensional and non-Abelian.

Both of the groups $\mathbb{R}^n$ and $SO(n)$ can be regarded as subgroups of the $n$-dimensional affine group $A(n)$, which is the semi-direct product $\mathbb{R}^n \lhd GL(n)$. The elements of $GL(n)$ that are not in $SO(n)$ are important to the motion of non-rigid bodies, so we briefly discuss the process of reducing from $GL(n)$ to $SO(n)$, although we shall have more to say about this in Part II.

As a first reduction, we restrict ourselves to orientation-preserving linear transformations. This means that one reduces to the identity component $GL^+(n)$ in $GL(n)$; the other connected component of $GL(n)$ is a diffeomorphic copy of $GL^+(n)$ that can be obtained by composing each of its elements with $-I$. The matrices of $GL^+(n)$ then all have positive determinant.

The next reduction is to the volume-preserving invertible transformations of $SL(n)$; the matrix of any such transformation will have unity determinant. In order to reduce any $A \in GL^+(n)$ to an element $\hat{A} \in SL(n)$, all that one needs to do is factor out the determinant:

$$\hat{A} = \det(A)^{-1/n} A \ . \tag{4.1}$$

One can then think of $GL^+(n)$ as a product group $\mathbb{R}^+ \times SL(n)$, where the multiplicative subgroup $\mathbb{R}^+$ is the *dilatation subgroup,* whose elements take the form $\lambda I$, where $\lambda \in \mathbb{R}^+$.

In order to reduce from $SL(n)$ to $SO(n)$, one needs to restrict to volume-preserving linear transformations that also preserve the Euclidian metric $\delta$; that is, if $R \in SO(n)$ then $R^T \delta R = \delta$. The factorization of $\hat{A}$ into a product $E_0 R$, where $E_0$ is a symmetric positive-definite matrix with unit determinant and $R \in SO(n)$ is by polar decomposition. Briefly, one sets [9]:

$$E_0 = \sqrt{\hat{A}\hat{A}^T} \ , \quad R = E_0^{-1}\hat{A} \ . \tag{4.2}$$

Since the manifold $E_0(n)$ of all $E_0$, which is not actually a subgroup of $SL(n)$, is diffeomorphic to $\mathbb{R}^{n(n+1)/2 \, - \, 1}$, we can then say that as a manifold $SL(n) = \mathbb{R}^{n(n+1)/2 \, - \, 1} \times SO(n)$. To summarize: we have shown that any $A \in GL(n)$ can be decomposed into a product $\pm \lambda E_0 R$, so as a manifold $GL(n) = \mathbb{Z}_2 \times \mathbb{R}^* \times \mathbb{R}^{n(n+1)/2 \, - \, 1} \times SO(n) = \mathbb{Z}_2 \times \mathbb{R}^{n(n+1)/2} \times SO(n)$.

One refers to $E_0 \in E_0(n)$ as a *finite strain.* Since the details of such transformations are more relevant to the study of the motion of deformable bodies, we shall return to that discussion in the next part of this series of articles.

Although the non-rigid motions of $(\mathbb{R}^n, \delta)$ seem to mostly describe motions of extended deformable bodies, it is still possible to consider point-like *pseudo-rigid* bodies [**25**]. Such bodies can then be described by a linear frame moving along a curve in the configuration manifold in such a manner that dilatations and strains of the frame are allowed. This is clearly a low-dimensional approximation to the motions of a deformable body, such as the motion of an elastic ball that is subject to dilatations, shears, rotations, and translations that are the same at every point of the body.

The main differences between the non-relativistic rigid motions and the relativistic ones stem from the fact that generally $n = 4$ in relativity and the metric is no longer the Euclidian metric, but the Minkowski one $\eta = \eta_{\mu\nu} dx^\mu dx^\nu$, where $\eta_{\mu\nu} = \mathrm{diag}(+1, -1, -1, -1)$. The orthogonal subgroup that has the most physical significance is then $SO(3, 1)$, which consists of orientation-preserving Lorentz transformations. Such transformations then satisfy the defining constraints:

$$\det A = 1, \qquad A^{\mathrm{T}} \eta A = \eta. \tag{4.3}$$

---

[9] For details of the proof that this prescription produces the desired result, see Chevalley [**24**].

As a consequence, the polar decomposition of an element of $GL(4)$ into an element of $E_0(n)$ and an element of $SO(4)$ is no longer as relevant, except insofar as it tells us what the topology of $GL(4)$ must be, irrespective of the choice of metric on $\mathbb{R}^4$.

One can still apply the reduction algorithm that was given above, except with the modification that now one sets:

$$E_0 = \sqrt{\hat{A}\hat{A}^*}, \qquad\qquad R = E_0^{-1}\hat{A}, \tag{4.4}$$

in which the * denotes the *Lorentz adjoint* of a matrix, namely, $A^* = \eta A^T \eta$. Hence, the basic property of Lorentz transformations then takes the form $A^{-1} = A^*$.

There is a possible snag in (4.4), due to the fact that since $\eta$ is not positive-definite, neither is $\hat{A}^T \eta \hat{A}$, and the matrix $E_0$ − hence, $R$ − might possibly be complex. However, one can show that both matrices are real by using the exponential map $\exp: \mathfrak{so}(3, 1) \rightarrow SO(3, 1)$. One expresses $E_0^2 = \hat{A}^T \eta \hat{A}$ as $\exp(2e_0)$, so one can set $E_0 = \exp(e_0)$.

Corresponding to the decomposition of $GL(n)$ into a product manifold that was given above, there is also a vector space decomposition of the Lie algebra $\mathfrak{gl}(n)$ into $\mathbb{R} \oplus \mathfrak{e}_0(n) \oplus \mathfrak{so}(n)$ that one obtains by polarizing an arbitrary $n \times n$ matrix $\alpha$ into the sum of a symmetric matrix $e$ and an anti-symmetric one $\omega$, and then subtracting off the trace of the symmetric matrix:

$$\varepsilon = \mathrm{Tr}(\alpha), \qquad e_0 = \tfrac{1}{2}(\alpha + \alpha^T) - \tfrac{1}{n}\varepsilon I, \qquad \omega = \tfrac{1}{2}(\alpha - \alpha^T). \tag{4.5}$$

The elements $\varepsilon \in \mathbb{R}$ then become the infinitesimal generators of dilatations, the elements $e_0 \in \mathfrak{e}_0(n)$ are the infinitesimal generators of strains, and the elements $\omega \in \mathfrak{so}(n)$ are the infinitesimal generators of Euclidian rotations.

For the Lorentz polarization, one uses the Lorentz adjoint instead of the transpose:

$$\varepsilon = \mathrm{Tr}(\alpha), \qquad e_0 = \tfrac{1}{2}(\alpha + \alpha^*) - \tfrac{1}{n}\varepsilon I, \qquad \omega = \tfrac{1}{2}(\alpha - \alpha^*). \tag{4.6}$$

The elements $\varepsilon \in \mathbb{R}$ are still the infinitesimal generators of dilatations, but the elements $e_0 \in \mathfrak{e}_0(3, 1)$ are the infinitesimal generators of Lorentz strains, and the elements $\omega \in \mathfrak{so}(3, 1)$ are the infinitesimal generators of Lorentz transformations.

## 5  Mechanical models for the motion of points

For the motion of a point in an $n$-dimensional manifold $M$, we shall use $K = [a, b]$, which represents a finite proper time interval. We shall regard all motions as the result of applying a $C^k$ one-parameter family – which is not necessarily a $C^k$ one-parameter subgroup, though – of elements of a group of motions $G$ to initial states, which take the form of points $x_0 \in U \subset M$, to produce $C^k$ curve segments in $M$. By prolongation, one derives an action of $J^k([a, b], G)$, that is, $C^k$ one-parameter families in $T^k G = G \times \mathfrak{g}[k]$, on the initial kinematical states of $J^k(K, U)$ to produce curve segments in $J^k(K, M)$.

### 5.1  Kinematical state spaces

It is most natural, from the standpoint of elementary physical mechanics, to regard the kinematical state of a moving point in a manifold $M$ as being defined by the position, velocity, and higher derivatives of a $C^k$ curve $\gamma: [a, b] \to M$, $\tau \mapsto x(\tau)$ for each value of the proper time parameter $\tau$. Hence, such a conception of a kinematical state implies that one is dealing with a section $\psi: [a, b] \to J^k([a, b], M)$ of the bundle of $k$-jets of curve segments in $M$. Its local form is then simply $\psi = (\tau, x^i(\tau), \dot{x}^i(\tau), \ldots, \overset{(k)}{x}{}^i(\tau))$. Since the manifold $[a, b]$ is contractible, the fibration $J^k([a, b], M) \to [a, b]$ is trivial and we can think of $J^k([a, b], M)$ as simply $[a, b] \times T^k(M)$.

One must be aware, of course, that past the first derivative the higher derivatives in a $k$-jet are purely local to each point and do not reflect the possible complex relationship between neighboring tangent spaces that necessitates the introduction of a connection on the tangent bundle to $M$ or the bundle of linear frames on $M$. It is possible to introduce connections within the context of jet bundles, but we shall not go into the details in the present study, except to point out their relationship to the integrability of the motion. (For the representation of connections in terms of jet bundles, see the discussion of jet fields in Saunders [**11**].)

Ultimately the highest order of differentiation in a kinematical state will be equal to the order of the dynamical equations. For instance, one notes that Newton's second law of motion defines a second-order system of differential equations, so the kinematical state $\psi$ terminates with the acceleration.

However, the *initial* kinematical state of such a system of equation will have an order that is one less than the order of the equations, as does the dynamical state, as we shall see. In order to represent an initial state $\psi_0 = (a, x_0^i, \ldots, \overset{(k)}{x}{}_0^i)$ as having the same order as the other states, one must accept that the remaining highest-order coordinates $\overset{(k)}{x}{}_0^i$ cannot be specified independently of the others, but must satisfy the constraint implied by the dynamical equations; that is, one must be starting with a solution of the system of dynamical equations.

Now, suppose one has a local action of $G$ on an open subset $U \subset M$ in the form of a smooth map $G \times U \to M$, $(g, x_0) \mapsto g x_0$. A motion of a point $x_0 \in U$ can also be defined by a $C^k$ curve segment $g: [a, b] \to G$, $\tau \mapsto g(\tau)$ that passes through the identity at $\tau = 0$. Its action on any point $x_0 \in U$ produces a curve $x(\tau) = g(\tau) x_0$ that takes points of $[a, b]$ to points in $M$. Hence, the orbits of the action of $g(\tau)$ on $U$ will define a congruence of curves in $M$.

By prolongation of the group action, one obtains an action $J^k([a, b], G) \times J^k([a, b], U)$ $\rightarrow J^k([a, b], M)$, $(\psi_G, \psi_0) \mapsto \psi$. Here, an *element* (not a section) $\psi_0 \in J^k([a, b], U)$ represents an initial kinematical state, so its local form is $(a, x_0^i, \dot{x}_0^i, \ldots, \overset{(k)}{x}_0^i)$. As for the section $\psi_G: [a, b] \rightarrow J^k([a, b], G)$, since we also have $J^k([a, b], G) = [a, b] \times T^k G = [a, b] \times G \times \mathfrak{g}[k]$ we can either represent it locally by $\psi_G = (\tau, g(\tau), \dot{g}(\tau), \ldots, \overset{(k)}{g}(\tau))$ or $(\tau, g(\tau), \omega(\tau), \overset{(1)}{\omega}(\tau), \ldots, \overset{(k-1)}{\omega}(\tau))$, in which the form that (3.23a, b) takes here is:

$$\omega = g^{-1}\dot{g}, \qquad \overset{(1)}{\omega} = g^{-1}\ddot{g} = \dot{\omega} + \omega\omega. \tag{5.1}$$

Further differentiations give all of the higher derivatives of $\omega$ in the form:

$$\overset{(k)}{\omega} = g^{-1}\overset{(k+1)}{g}, \tag{5.2}$$

which can also be expressed as:

$$\overset{(k+1)}{g} = g\,\overset{(k)}{\omega}. \tag{5.3}$$

We can introduce a differential operator that behaves like a covariant derivative operator on maps $[a, b] \rightarrow \mathfrak{g}[k]$, namely:

$$\nabla \overset{(l)}{\omega} = \frac{d\,\overset{(l)}{\omega}}{d\tau} + \omega\,\overset{(l)}{\omega}, \qquad l = 1, \ldots, k\,. \tag{5.4}$$

This then makes the recursion (5.2) take the form:

$$\overset{(l+1)}{\omega} = \nabla\overset{(l)}{\omega}, \qquad l = 1, \ldots, k-1. \tag{5.5}$$

The action of a section $\psi_G: [a, b] \rightarrow J^k([a, b], G)$ on an initial state $\psi_0 \in J^k([a, b], U)$ can then be obtained in local form by specializing (3.20a, b, c) and (3.22a, b, c). In the former case, we get:

$$x = gx_0, \qquad \dot{x} = \dot{g}x_0 + g\dot{x}_0, \quad \ddot{x} = \ddot{g}x_0 + 2\dot{g}\dot{x}_0 + g\ddot{x}_0, \quad \ldots, \tag{5.6}$$

and in the latter:

$$x = gx_0, \qquad \dot{x} = g(\omega x_0 + \dot{x}_0), \qquad \ddot{x} = g(\overset{(1)}{\omega}x_0 + 2\omega\dot{x}_0 + \ddot{x}_0), \qquad \ldots \tag{5.7}$$

We can define our $\nabla$ operator on sections of $J^k([a, b], U)$ in the obvious way:

$$\nabla \overset{(l)}{x}_0 = \omega \overset{(l)}{x}_0 + \overset{(l+1)}{x}_0 \, , \qquad l = 1, \dots, k, \tag{5.8}$$

which then puts (5.7) into the form:

$$x = gx_0, \qquad \dot{x} = g\nabla x_0, \qquad \ddot{x} = g\nabla^2 x_0 \, , \qquad \dots, \qquad \overset{(k)}{x} = g\nabla^k x_0 \, , \tag{5.9}$$

since the $\nabla$ operator is a linear derivation on the sections.

### 5.2  The integrability of kinematical states

So far, we have defined our sections of the bundles $J^k([a, b]; M) \to [a, b]$ and $J^k([a, b], G) \to [a, b]$ by starting with a curve in $M$ or $G$, respectively, and going to successively higher derivatives; i.e., the section is the $k^{\text{th}}$ prolongation of the curve. However, as we pointed out above, not all sections of these bundles can be represented as prolongations of curves, but only the integrable sections.

When we represent a kinematical state as a section $\psi$ of $J^k([a, b]; M) \to [a, b]$, the integrability condition is simply:

$$D\psi = 0. \tag{5.10}$$

in which $D: J^k([a, b]; M) \to T^*([a, b]) \otimes J^{k-1}([a, b]; M)$ is the Spencer operator, which takes the following form here:

$$D\psi = \left( \tau, Dx^i, D\dot{x}^i, \cdots, D \overset{(k-1)}{x}{}^i \right) d\tau \, , \tag{5.11}$$

with:

$$Dx^i = \frac{dx^i}{d\tau} - \overset{(1)}{x}{}^i \, , \qquad \dots, D \overset{(k-1)}{x}{}^i = \frac{d \overset{(k-1)}{x}{}^i}{d\tau} - \overset{(k)}{x}{}^i \, . \tag{5.12}$$

Hence, in order for a general section $\psi(\tau) = (\tau, x^i(\tau), \overset{(1)}{x}{}^i(\tau), \dots \overset{(k)}{x}{}^i(\tau))$ to be integrable, i.e., the prolongation of a curve $\chi(\tau)$, one must have:

$$\overset{(1)}{x}{}^i(\tau) = \frac{dx^i}{d\tau} \, , \dots, \overset{(k)}{x}{}^i(\tau) = \frac{d \overset{(k-1)}{x}{}^i}{d\tau} \, , \tag{5.13}$$

i.e., each successive set of components must be the proper time derivative of the previous set.

Note that since $\dim([a, b]) = 1$ we must have $\Lambda^2([a, b]) = 0$ and the Spencer sequence terminates after the first application of $D$. This has the effect of implying that all sections of $T^*([a, b]) \otimes J^{k-1}([a, b]; M) \to [a, b]$ must be integrable.

In order to find the integrability condition for a section of the bundle $J^k([a, b]; G) \times J^k([a, b]; U) \to [a, b]$, when one represents it as $\psi(\tau) = (\psi_G(\tau), \psi_0) = (\tau, g(\tau), \overset{(1)}{g}(\tau),$ $\ldots, \overset{(k)}{g}(\tau)) \times (a, x_0, \dot{x}_0, \cdots, \overset{(k)}{x}_0)$, we must first examine the form that the Spencer operator takes on sections of that bundle.

The Spencer operator in this case takes the form $D$: $J^k([a, b]; G) \times J^k([a, b]; U) \to T^*([a, b]) \otimes (J^{k-1}([a, b]; G) \times (J^{k-1}([a, b]; U))$, in such a way that we can say that:

$$D(\psi_G, \psi_U) = (D\psi_G, D\psi_U), \tag{5.14}$$

with:

$$D\psi_G = \left(\tau, Dg_j^i, D\dot{g}_j^i, \cdots, D \overset{(k-1)}{g}_j^i\right) d\tau, \tag{5.15a}$$

$$D\psi_U = \left(a, Dx_0^i, D\dot{x}_0^i, \cdots, D \overset{(k-1)}{x}_0^i\right) d\tau, \tag{5.15b}$$

in which:

$$D \overset{(l)}{g}_j^i = \frac{d \overset{(l)}{g}_j^i}{d\tau} - \overset{(l+1)}{g}_j^i, \qquad D \overset{(l)}{x}_0^i = - \overset{(l+1)}{x}_0^i, \quad l = 1, \ldots, k-1. \tag{5.16}$$

Note that even though the initial kinematical state $\psi_0$ is not a section, and thus does not vary in time, the Spencer operator still acts on it. However, although saying that a section of the bundle $J^k([a, b]; G)$ is integrable is equivalent to saying that $D\psi_G = 0$, the condition $D\psi_U = 0$ is satisfied only for initial states with $\overset{(l)}{x}_0 = 0$ for $l > 0$, which amounts to an initial state of rest. This leads to an important difference between the integrability of kinematical states in the latter bundle, which only implies that the successive terms in the group state are successive derivatives, and the integrability of kinematical states in $J^k([a, b]; M)$.

In order to see this, one must relate $(D\psi_G, D\psi_U)$ to $D\psi$ by means of the group action. If one expresses the relationship between $\psi$ and $(\psi_G, \psi_U)$ in the form $\psi = \psi_G \cdot \psi_U$ the one finds that the relationship between $D\psi$ and $(D\psi_G, D\psi_U)$ can be expressed in the form:

$$D\psi = D\psi_G \cdot \psi_U + \psi_G \cdot D\psi_U. \tag{5.17}$$

In order to find the coordinate form for this, one uses the rules given in (5.6) to deduce that:

$$Dx^i = Dg_j^i x_0^j + g_j^i Dx_0^j = Dg_j^i x_0^j - g_j^i \dot{x}_0^j \tag{5.18a}$$
$$D\dot{x}^i = [Dg_j^i \dot{x}_0^j + D\dot{g}_j^i x_0^j] + [\dot{g}_j^i Dx_0^j + g_j^i D\dot{x}_0^j],$$

$$=[Dg^i_j \dot{x}^j_0 + D\dot{g}^i_j x^j_0] - [\dot{g}^i_j \dot{x}^j_0 + g^i_j \ddot{x}^j_0] \tag{5.18b}$$

In (5.18a, b), we have grouped the terms according to whether they involve coordinates in $J^k([a, b]; G)$ or $J^k([a, b]; U)$. We have also stopped at $k = 2$, since clearly the general expression would be quite cumbersome to specify, although it also clearly derives from the binomial expansion.

It is important to see how (5.17) implies that the integrability of a kinematical state in $J^k([a, b]; M)$ does not have to be equivalent to the integrability of a state in the form of $(\psi_G, \psi_U)$. In particular, if $\psi$ is integrable, so $D\psi = 0$, then this implies the condition:

$$0 = D\psi_G \cdot \psi_U + \psi_G \cdot D\psi_U, \tag{5.19}$$

which does not have to imply $D\psi_G = 0$, since $D\psi_U$ will not vanish except for certain initial states. Conversely, one can see that if $\psi_G$ is integrable as a section of $J^k([a, b]; G)$ then for an arbitrary initial state $\psi_U$ the resulting state $\psi_G \cdot \psi_U$ does not have to be integrable, either.

Locally, if $D\psi = 0$ then, to second order, $Dx^i = D\dot{x}^i = 0$, and we see that if the initial state $\psi_U = (a, x^i_0, \dot{x}^i_0)$ is arbitrary then the resulting condition on $\psi_G = (\tau, g^i_j(\tau), \dot{g}^i_j(\tau))$ is:

$$Dg^i_j x^j_0 = g^i_j \dot{x}^j_0, \qquad Dg^i_j \dot{x}^j_0 + D\dot{g}^i_j x^j_0 = \dot{g}^i_j \dot{x}^j_0 + g^i_j \ddot{x}^j_0. \tag{5.20}$$

We can write them out explicitly as:

$$\frac{dg^i_j}{d\tau} x^j_0 = \dot{g}^i_j x^j_0 + g^i_j \dot{x}^j_0, \tag{5.21a}$$

$$\frac{d}{d\tau}(\dot{g}^i_j x^j_0 + g^i_j \dot{x}^j_0) = \ddot{g}^i_j x^j_0 + 2\dot{g}^i_j \dot{x}^j_0 + g^i_j \ddot{x}^j_0, \tag{5.21b}$$

which are seen to follow from differentiation of the basic relations (5.6).

Conversely, if one applies an integrable state $\psi_G$ in $J^k([a, b]; G)$ to an arbitrary initial state $\psi_U$ in $J^k([a, b]; U)$ then the resulting state $\psi$ in $J^k([a, b]; M)$ will have the property that:

$$Dx^i = -g^i_j \dot{x}^j_0, \qquad D\dot{x}^i = -\dot{g}^i_j \dot{x}^j_0 - g^i_j \ddot{x}^j_0. \tag{5.22}$$

Clearly, a general initial state will not produce an integrable state under the action of the group of motions.

If we wish to work with the kinematical state in $J^k([a, b]; G)$ in the form $(\tau, g, \omega, \overset{(1)}{\omega}, ..., \overset{(k-1)}{\omega})$ then we need to alter the form of $D\psi_G$ accordingly. Now we should have:

$$D\psi_G = \left(\tau, Dg^i_j, D\omega^i_j, \cdots, D\overset{(k-1)}{\omega}{}^i_j\right)d\tau \,. \tag{5.23}$$

We can then use the rules (5.3) to relate the components of (5.23) to those of (5.17a):

$$Dg^i_j = \frac{dg^i_j}{d\tau} - \dot{g}^i_j = \frac{dg^i_j}{d\tau} - g^i_k\omega^k_j \,, \tag{5.24a}$$

$$D\dot{g}^i_j = \frac{d(g^i_k\omega^k_j)}{d\tau} - \ddot{g}^i_j = (Dg^i_k)\omega^k_j + g^i_k D\omega^k_j \,. \tag{5.24b}$$

These expressions show that the integrability of a section of $J^k([a, b]; G)$ in one form is equivalent to the integrability in the other form.

Substituting (5.24b) in (5.18b) then gives:

$$D\dot{x}^i = Dg^i_j\nabla x^j_0 + g^i_k D\omega^k_j x^j_0 - g^i_j\nabla \dot{x}^j_0 \,. \tag{5.25}$$

Hence, if $\psi$ is integrable then we must have:

$$Dg^i_j x^j_0 = g^i_j \dot{x}^j_0 \,, \qquad Dg^i_j\nabla x^j_0 + g^i_k D\omega^k_j x^j_0 = g^i_j\nabla \dot{x}^j_0 \,. \tag{5.26}$$

and if $\psi_G$ is integrable then we must have:

$$D\dot{x}^i = -g^i_j \dot{x}^j_0 \,, \qquad D\dot{x}^i = -g^i_j\nabla \dot{x}^j_0 \,. \tag{5.27}$$

Explicitly equations (5.26) take the form:

$$\frac{dg^i_j}{d\tau} x^j_0 = g^i_j\nabla x^j_0 \,, \qquad \frac{d}{d\tau}(g^i_j\nabla x^j_0) = g^i_j\nabla^2 x^j_0 \,, \tag{5.28}$$

which are equivalent to (5.22a, b), and are seen to follow from the differentiation of (5.9).

We also note that from (5.4) one has:

$$D\omega^i_j = \frac{d\omega^i_j}{d\tau} - \dot{\omega}^i_j = \nabla\omega^i_j - \overset{(1)}{\omega}{}^i_j \,. \tag{5.29}$$

and, in fact:

$$D\overset{(l)}{\omega}{}^i_j = \nabla\overset{(l)}{\omega}{}^i_j - \overset{(l+1)}{\omega}{}^i_j \,, \qquad l = 0, \ldots, k-2. \tag{5.30}$$

Therefore, the action of the Spencer operator on sections of $J^k([a, b]; G)$, when expressed locally in the present form, can also be related to the action of the $\nabla$ operator.

One must naturally wonder whether the constraint of integrability is necessary in order to be describing physically realizable motions, since one is first introduced to integrable kinematical states in mechanics. In order to see that non-integrable motions are physically realizable, one need only recall the example given above in which an integrable $k$-jet $\psi_G = (\tau, g(\tau), \ldots, \overset{(k)}{g}(\tau)) \in J^k([a, b], G)$ is applied to an arbitrary initial kinematical state $\psi_U = (a, x_0, \dot{x}_0, \cdots, \overset{(k)}{x}_0) \in J^k([a, b], U)$ and produces a non-integrable section of $J^k([a, b], G) \to [a, b]$.

One can also consider the case where acceleration must be defined by the covariant derivative of velocity in the direction of motion using a connection on the tangent bundle $T(M)$, which we represent by a matrix of $\omega^i_j$ of 1-forms relative to a natural frame field. A kinematical state is then represented by a 2-jet of the form $\psi(\tau) = (\tau, x^i(\tau), \dot{x}^i(\tau), a^i(\tau))$ with:

$$\dot{x}^i(\tau) = \frac{dx^i}{d\tau}, \quad a^i(\tau) = \frac{d\dot{x}^i}{d\tau} + \omega^i_j(x, \dot{x})\dot{x}^j. \tag{5.31}$$

This makes:

$$D\psi = -\omega^i_j(x, \dot{x})\dot{x}^j \frac{\partial}{\partial \dot{x}^i} \otimes d\tau, \tag{5.32}$$

which is generally non-zero, even in the case of geodesic motion, for which $a^i(\tau) = 0$.

A further indication that integrability is not always physically necessary is given by the example of motion with anholonomic constraints. An anholonomic constraint on a configuration manifold $M$ is a non-integrable sub-bundle of its tangent bundle. That is, there is no foliation of $M$ by integral submanifolds, whose tangent spaces are then, by definition, equal to the fibers of the sub-bundle. The most common example of such a constraint is that of a disc rolling without slipping on a plane.

### 5.3 Dynamical state spaces

As a motivation for our definition of a dynamical state, we start with an action functional:

$$S[x(\tau)] = \int_a^b \mathcal{L}(\psi)d\tau = \int_a^b \mathcal{L}(\tau, x^i(\tau), \dot{x}^i(\tau), \cdots, \overset{(k)}{x}^i(\tau))d\tau, \tag{5.33}$$

in which $\mathcal{L}: J^k([a, b], M) \to \mathbb{R}$ is a smooth function on the kinematical state space that represents a Lagrangian function for the mechanical system in question.

By exterior differentiating the integrand, we get:

$$(d\mathcal{L} \wedge d\tau)|_\psi = \left(\frac{\partial \mathcal{L}}{\partial \tau}d\tau + \frac{\partial \mathcal{L}}{\partial x^i}dx^i + \frac{\partial \mathcal{L}}{\partial \dot{x}^i}d\dot{x}^i + \cdots + \frac{\partial \mathcal{L}}{\partial \overset{(k)}{x}^i}d\overset{(k)}{x}^i\right) \wedge d\tau, \tag{5.34}$$

which we suggestively rewrite in the form:

$$(d\mathcal{L} \wedge d\tau)|_\psi = \left( F_i dx^i + p_i d\dot{x}^i + \cdots + \overset{(k)}{p_i} d \overset{(k)}{x^i} \right) \wedge d\tau \tag{5.35}$$

Note that the first term in $d\mathcal{L}$ disappears when one takes the exterior product $d\mathcal{L} \wedge d\tau$. Hence, it does not appear in the Euler-Lagrange equations that follow from (5.33), so we do not include it in our ultimate dynamical state.

Since $J^k([a, b], M) = [a, b] \times T^k(M)$, we can unambiguously project $d\mathcal{L}$ from a 1-form on $J^k([a, b], M)$ to a 1-form on $T^k(M)$. With a slight generalization, we then define our dynamical states to be elements $\phi \in T([a, b]) \otimes V^*(J^{k-1}([a, b], M))$, which then take the local form:

$$\phi = (F_i dx^i + p_i d\dot{x}^i + \cdots + \overset{(k-2)}{p_i} d\overset{(k-1)}{x^i}) \otimes \frac{d}{d\tau} \tag{5.36}$$

Our justification for this generalization is simply that dynamical states of this form are 1-forms on the images $D\psi$ of the Spencer operator when it acts on kinematical states, and we shall use such expressions in the next section to characterize the dynamical laws.

The first term in the parentheses in (5.36) appears to represent a differential increment of *work* along the path, so the components $F_i$ represent a *generalized force*. The second term represents a differential increment of *kinetic energy* and its components $p_i$ represent a *generalized momentum*.

One can think of the components $\overset{(m)}{p_i}$ as representing successive proper time integrals of the preceding components $\overset{(m-1)}{p_i}$, instead of successive time derivatives, since all of the terms in the parentheses in (5.36) must have the unit of energy and the units of the differentials $d \overset{(m)}{x^i}$ are increasing in powers of $1/\tau$. Since most of physical mechanics is based on second-order equations there are no widely-discussed physical interpretations for the terms $\overset{(m)}{p_i}$ for $m > 1$. However, one does note that $\overset{(1)}{p_i}$ has the units of a mass moment, such as one uses in the computation of the center of mass. The coupling of mass moment with acceleration to give a form of energy is not discussed in conventional mechanics, though.

It is also important to understand that the fact that $\partial\mathcal{L}/\partial\tau$ plays no role in the ultimate equations of motion represents one limitation of the Lagrangian (and Hamiltonian) methodology: it is inapplicable to the case of time-varying Lagrangians, such as one encounters with dissipative systems. Furthermore, the fact that we are obtaining our basic dynamical objects from the components of an exact 1-form, namely $d\mathcal{L}$, means that the components of a more general 1-form would not represent forces and momenta that are associated with a variational problem.

Strictly speaking, the bundle $\Lambda^1(T^k(M))$ is not dual to our kinematical state space $J^k([a, b]; M) = [a, b] \times T^k(M)$, but to the vertical, i.e., time-independent, part of its tangent bundle $T(J^{k-1}([a, b]; M))$, namely, $T(T^{k-1}(M))$. This amounts to the statement that dynamical states represent responses to infinitesimal changes – i.e., variations – in the kinematical state, not the state itself.

Furthermore, we are only using 1-forms on $J^{k-1}([a, b]; M) = [a, b] \times T^{k-1}(M)$. This implies that the dimension of the space of dynamical states is lower than the dimension of the space of kinematical states. Hence, the dimension of the space of dynamical states equals the dimension of the space of initial states.

Now, suppose that the kinematical state $\psi$ is the result $\psi_G \cdot \psi_0$ of the non-uniform action $G \times U \to M$ of a group $G$ on $U \subset M$, which is then prolonged to an action $J^k([a, b], G) \times J^k([a, b], U) \to J^k([a, b], M)$.

A 1-form $\phi \in T([a, b]) \otimes V^*(J^{k-1}([a, b], M)$ then pulls back to a 1-form in $T([a, b]) \otimes [V^*(J^{k-1}([a, b], G) \times J^{k-1}([a, b], U))]$, and since the latter bundle decomposes into a Whitney sum $[T([a, b]) \otimes V^*(J^{k-1}([a, b], G))] \oplus [T([a, b]) \otimes V^*(J^{k-1}([a, b], U))]$, any 1-form $\phi$ in that bundle can be expressed uniquely as a sum $\phi_G + \phi_U$, in which the term $\phi_U$ represents the initial value of the dynamical state, whereas the term $\phi_G$ represents the time-varying part. Furthermore, one can take advantage of the fact that $J^{k-1}([a, b], G) = [a, b] \times T^{k-1}(G)$, so $V^*(J^{k-1}([a, b], G)) = (T^{k-1})^*(G)$.

From (5.4), we substitute:

$$dx^i = dg^i_j x^j_0 + g^i_j dx^j_0, \tag{5.37a}$$

$$d\dot{x}^i = d\dot{g}^i_k x^k_0 + \dot{g}^i_j dx^j_0 + dg^i_j \dot{x}^j_0 + g^i_j d\dot{x}^j_0 , \tag{5.37b}$$

in (5.35), and we get, if $k = 2$:

$$\phi_U = \left[ \tilde{F}_{0i} dx^i_0 + p_{0i} d\dot{x}^i_0 \right] \otimes \frac{d}{d\tau} , \tag{5.38a}$$

$$\phi_G = \left[ \tilde{T}^j_i dg^i_j + L^j_i d\dot{g}^i_j \right] \otimes \frac{d}{d\tau} , \tag{5.38b}$$

into which we have introduced the notations:

$$\tilde{F}_{0i} = F_{0i} + p_j \dot{g}^j_i , \qquad F_{0i} = F_j g^j_i , \qquad p_{0i} = p_j g^j_i , \tag{5.39a}$$

$$\tilde{T}^j_i = T^j_i + p_i \dot{x}^j_0 , \qquad T^j_i = F_i x^j_0 , \qquad L^j_i = p_i x^j_0 . \tag{5.39b}$$

The expressions $F_{0i}$ and $p_{0i}$ then represent the components of the 1-forms $F$ and $p$ pulled back to the initial state by means of the group action. However, one should be careful about calling them the initial values of $F_i$ and $p_i$ when the function $g(\tau)$ is non-constant, since the derivative of the group action would not generally coincide with the group action in that case, and the pull-backs of $F_i$ and $p_i$ by the group action would differ from $F_{0i}$ and $p_{0i}$. The expressions $T^j_i$ and $L^j_i$ then represent the generalized torque and angular momentum of the motion, as measured in an inertial frame.

When one chooses to represent $J^{k-1}([a, b], G)$ as $[a, b] \times G \times \mathfrak{g}[k]$ instead by left-translation $g \in G$ to the identity, the bundle $V^*(J^{k-1}([a, b], G))$ takes the form $T^*(G) \times \mathfrak{g}^*[k]$.

We first compute:

$$d\dot{g}^i_j = dg^i_k \omega^k_j + g^i_k d\omega^k_j . \tag{5.40}$$

and then replace $dg$ with $g\theta$, where $\theta$ is the *Maurer-Cartan* 1-form for $G$:

$$\theta^i_j = \tilde{g}^i_k dg^k_j ; \tag{5.41}$$

the tilde on the $g$ indicates the inverse of the matrix.

The Maurer-Cartan 1-form is not exact, or even closed, in general. Rather, it satisfies the *Maurer-Cartan* equations:

$$d\theta^a = -\tfrac{1}{2} c^a_{bc} \theta^b \wedge \theta^c , \tag{5.43}$$

in which the $c$'s are the structure constants for the Lie algebra $\mathfrak{g}$ and we have temporarily given the basis elements for $\mathfrak{g}$ one index that ranges from 1 to $\dim(\mathfrak{g})$, instead of the two matrix indices. These equations express either the vanishing of the curvature of the 1-form $\theta$, when regarded as a connection on the (trivial) bundle of $G$ frames in $T(G)$, or the complete integrability of the exterior differential system $\theta = 0$, whose integral submanifolds will be of dimension zero; viz., the points of $G$.

For an integrable section of $J^k([a, b], G)$, one will have $dg = g\omega d\tau$, which then gives:

$$\theta^i_j = \omega^i_j d\tau . \tag{5.44}$$

We then have:

$$dg^i_j = g^i_k \theta^k_j , \qquad d\dot{g}^i_j = g^i_k (\theta^k_l \omega^l_j + d\omega^k_j) . \tag{5.45}$$

We can represent the dynamical state $\phi$ in the form:

$$\phi_G = \left[ \tilde{\mathcal{T}}_i{}^j \theta^i_j + \mathcal{L}^j_i d\omega^i_j \right] \otimes \frac{d}{d\tau} , \tag{5.46a}$$

$$\phi_U = \left[ \tilde{F}_{0i} dx^i_0 + p_{0i} d\dot{x}^i_0 \right] \otimes \frac{d}{d\tau} , \tag{5.46b}$$

and this time we have introduced the notations:

$$\tilde{\mathcal{T}}_i{}^j = \mathcal{T}_i{}^j + \mathcal{L}^k_i \omega^j_k + p_{0i} \dot{x}^j_0 , \qquad \mathcal{T}_i{}^j = F_{0i} x^j_0 , \qquad \mathcal{L}^j_i = p_{0i} x^j_0 . \tag{5.47a}$$

$$\tilde{F}_{0i} = F_{0i} + p_{0j}\omega_i^j\,. \tag{5.47b}$$

We see that the expressions $\mathcal{T}_i^{\ j}$ then define a 1-form $\mathcal{T} = \mathcal{T}_i^{\ j}\theta_j^i$ on $G$ that represents a differential increment of work done by the generalized torque, and the $\mathcal{L}_i^j$ define a 1-form $\mathcal{L} = \mathcal{L}_i^j d\omega_j^i$ that represents a differential increment of kinetic energy due to the generalized angular momentum. With the replacements (5.45), all of the dynamical variables in (5.47a, b) are then expressed in the rest frame of $\phi_U$. We shall see in the examples that this is a generalization of the usual rotating frame that one would define in rotational mechanics.

Something that has been hidden by our restriction to $k = 2$ is the fact that had we gone to $k = 3$, we would have also made additional contributions to *all* of the terms in both $\phi_U$ and $\phi_G$, and not just the next order terms, since:

$$d\ddot{x}^i = (\ddot{g}_j^i dx_0^j + 2\dot{g}_j^i d\dot{x}_0^j + g_j^i d\ddot{x}_0^j) + (d\ddot{g}_j^i x_0^j + 2d\dot{g}_j^i \dot{x}_0^j + dg_j^i \ddot{x}_0^j))\,. \tag{5.48}$$

Similarly, each further differentiation of $x(\tau)$ will add terms to all of the lower-level terms in $\phi_U$ and $\phi_G$. Hence, although one can safely truncate the order of differentiation in any mechanical model that is based on kinematical states in $J^k([a, b], M)$, we see that, in principle, things are not so simple when we deal with kinematical states in $J^k([a, b], G)$. When we discuss continuum mechanics, we shall see that the issue becomes one of the type of $\mathfrak{g}$, and that only when $\mathfrak{g}$ has finite type can one truncate without consequence; even then, one might have a higher type than $k = 2$ would suggest.

We then express our dynamical state as a 1-form $\phi \in T([a, b]) \otimes V^*(J^{k-1}([a, b], G)) = T([a, b]) \times T^{k-1}(G) = T([a, b]) \times G \times \mathfrak{g}[k{-}1])$, whose components take the form $(\mathcal{T}_i^{\ j}, \mathcal{L}_i^j, \cdots, \overset{(k-2)}{\mathcal{L}_i^j})$. Hence, they can be evaluated on vectors tangent to $J^{k-1}([a, b]; G)$, which again represent variations of kinematical states.

If we wish to derive $\phi$ from a Lagrangian in the case of a group action then we can pull the Lagrangian on $J^k([a, b], M)$ back to a function on $J^k([a, b], G) \times J^k([a, b], U)$ by using the group action, and if we choose an initial kinematical state $\psi_0 = (\tau_0, x_0, \dot{x}_0, \cdots)$ then we can regard $\mathcal{L}$ as a function on $J^k([a, b], G)$.

### 5.4  Integrability of dynamical states

In order to motivate the definition of the dual of the Spencer operator that acts on dynamical states, we start with the vertical 1-form part of a general second order dynamical state, namely:

$$\phi = F_i \, dx^i + \ p_i d\dot{x}^i\,, \tag{5.49}$$

and pull it back from $J^2([a, b], M)$ to $[a, b]$ along a general section $\psi : [a, b] \rightarrow J^2([a, b], M)$, with the local form $\psi = (\tau, x^i(\tau), \dot{x}^i(\tau), \ddot{x}^i(\tau))$. However, if we do not assume that $\psi$ is integrable then we can effect this pull-back by the following replacements:

$$dx^i = (\dot{x}^i + Dx^i)\, d\tau, \qquad d\dot{x}^i = (\ddot{x}^i + D\dot{x}^i)\, d\tau. \tag{5.50}$$

This makes:

$$\psi^* \phi = (F_i Dx^i + p_i D\dot{x}^i + F_i \dot{x}^i + p_i \ddot{x}^i)\, d\tau. \tag{5.51}$$

By an application of the product rule for differentiation, this takes the form:

$$\psi^* \phi = \left[ F_i Dx^i + p_i D\dot{x}^i + \left( F_i - \frac{dp_i}{d\tau} \right) \dot{x}^i + \frac{d(p_i \dot{x}^i)}{d\tau} \right] d\tau, \tag{5.52}$$

which we then put into the form:

$$\psi^* \phi = \left[ F_i Dx^i + p_i D\dot{x}^i + (D^* F_i) \dot{x}^i + (D^* p_i) \ddot{x}^i + \frac{d(p_i \dot{x}^i)}{d\tau} \right] d\tau, \tag{5.53}$$

by defining:

$$D^* F_i = F_i - \frac{dp_i}{d\tau}, \qquad D^* p_i = 0. \tag{5.54}$$

One notes that in the event that $\phi = d\mathcal{L}$ for some Lagrangian $\mathcal{L}$ on $J^2([a, b], M)$ the resulting expression for $D^* F_i$ is identical with $\delta \mathcal{L}/dx^i$. However, the sequences of calculations that we made are valid for more general 1-forms on $J^2([a, b], M)$ than just exact ones. One must also note that $\ddot{x}^i$ no longer figures explicitly in either (5.52) or (5.53), as it has been absorbed into $D\dot{x}^i$ and it gets multiplied by zero in (5.53)

If we further define the vertical 1-form on $J^2([a, b], M)$:

$$D^* \phi = (D^* F_i) dx^i + (D^* p_i) d\dot{x}^i, \tag{5.55}$$

with the replacements (5.54), then its pullback to $[a, b]$ along an integrable $\psi$ is:

$$\psi^* D^* \phi = [(D^* F_i) \dot{x}^i + (D^* p_i) \ddot{x}^i] d\tau, \tag{5.56}$$

and (5.53) can be put into the form:

$$\psi^* \phi = \left[ \phi(D\psi) + D^* \phi(\psi) + \frac{d(p_i \dot{x}^i)}{d\tau} \right] d\tau. \tag{5.57}$$

Furthermore, since $d[\phi(\psi)]/d\tau$ differs from $d(p_i\dot{x}^i)/d\tau$ only by a time derivative, we amend this to say:

$$\psi^*\phi = \left[\phi(D\psi) + D^*\phi(\psi) + \frac{d}{d\tau}\phi(\psi)\right]d\tau \,. \tag{5.58}$$

We then state our generalization of the least-action principle in the form:

**Theorem:**

*a. If $\psi$ is an integrable kinematical state and $\phi$ is a dynamical state then $\psi^*\phi$ is an exact 1-form on $[a, b]$ iff $D^*\phi(\psi)$ is a time derivative.*

*b. A sufficient condition for $\psi^*\phi$ to be exact for any integrable section $\psi$ is that:*

$$D^*\phi = 0 \,. \tag{5.59}$$

Hence, we say that $\phi$ is *weakly integrable* iff $D^*\phi(\psi)$ is a time derivative whenever $\psi$ is any integrable section and *strongly integrable* iff $D^*\phi = 0$. (Here, we are using the term "strong" to mean "less solutions.")

We then define the operator $D^*: T([a, b]) \otimes V^*(J^{k-1}([a, b], M)) \to V^*(J^k([a, b], M))$ to be the pull-back (i.e., transpose or adjoint) of the Spencer operator $D: J^k([a, b], M) \to T^*[a, b] \otimes J^{k-1}([a, b], M)$, namely, if $\phi \in T([a, b]) \otimes V^*(J^{k-1}([a, b], M))$ then:

$$D^*\phi|_\psi = -\phi|_{D\psi} - \frac{d}{d\tau}\phi\Big|_\psi \,. \tag{5.60}$$

The reader will find that this construction is a specialization of one that was proposed in Pommaret [**1**].

If we examine the resulting equations (5.59) to second order then we see that locally the non-trivial ones take the form:

$$F_i = \frac{dp_i}{d\tau} \,, \tag{5.61}$$

which is either the form of Newton's second law or the Euler-Lagrange equations when $\phi$ is based in an exact form. Had we gone to the next order in $\phi$, we would have seen that the integrability condition would also tell us that momentum is the proper time derivative of the mass moment.

Hence, we postulate that equation (5.59) is a reasonable generalization of Newton's second law. We also see that the Euler-Lagrange equations that follow from the usual fixed-endpoint assumption concerning variations of a curve $\gamma$ and an integration by parts, to second order, are also related to (5.59), which is also consistent with our postulate. However, since not all mechanical models can be put into Lagrangian form, we see that (5.59) is broader in scope than the Euler-Lagrange equations.

Now, let us look at the representation of dynamical states as vertical 1-forms $\phi_G + \phi_U$ on $T([a, b]) \otimes J^{k-1}([a, b], G) \times J^{k-1}([a, b], U)$, such as in (5.38a, b).

Since $D\psi$ takes the form (5.19), we then have:

$$\phi(D\psi) = \phi(D\psi_G \cdot \psi_U) + \phi(\psi_G \cdot D\psi_U) \,. \tag{5.62}$$

We then expect $D^*$ to behave like:

$$D^*\phi(\psi) = D^*\phi(\psi_G) \cdot \psi_U + \psi_G \cdot D^*\phi(\psi_U) \tag{5.63}$$

which we rewrite in the form:

$$D^*\phi(\psi) = D^*\phi_G(\psi_G) + D^*\phi_U(\psi_U) \tag{5.64}$$

in which we formally define $D^*\phi_G$ and $D^*\phi_U$ as:

$$D^*\phi_G = D^*\phi \otimes \psi_U \,, \qquad D^*\phi_U = \psi_G \cdot D^*\phi \,. \tag{5.65}$$

In order to clarify the meaning of these formal definitions, we express $D^*\phi_G$ and $D^*\phi_U$ in the local form:

$$D^*\phi_G = (D^*\tilde{T}_i^{\ j})dg_j^i + (D^*L_i^j)d\dot{g}_j^i \,, \tag{5.66a}$$

$$D^*\phi_U = (D^*\tilde{F}_{0i})dx_0^i + (D^*p_{0i})d\dot{x}_0^i \,. \tag{5.66a}$$

If we reason by analogy with the formulas (5.54):

$$D^*\tilde{T}_i^{\ j} = \tilde{T}_i^{\ j} - \frac{dL_i^j}{d\tau} \,, \qquad D^*L_i^j = 0 \,, \tag{5.67a}$$

$$D^*\tilde{F}_{0i} = \tilde{F}_{0i} - \frac{dp_{0i}}{d\tau} \,, \qquad D^*p_{0i} = 0 \,. \tag{5.67b}$$

Note that although the differentials $dx_0^i$ and $d\dot{x}_0^i$ are not time-varying, and therefore do not contribute a time derivative to either $D(dx_0^i)$ or $D(d\dot{x}_0^i)$, nevertheless, since $\tilde{F}_{0i}$ and $p_{0i}$ are time-varying they do contribute time derivatives to the adjoint of $D$ in both cases.

The equations that one derives from (5.57) are then:

$$\tilde{T}_i^{\ j} = \frac{dL_i^j}{d\tau} \,, \qquad\qquad \tilde{F}_{0i} = \frac{dp_{0i}}{d\tau} \,. \tag{5.68}$$

If we go back to the definitions (5.39a, b) of $\tilde{T}_i^{\ j}$ and $F_{0i}$ then we see that these equations take the form:

$$T_i^{\ j} = \frac{dL_i^j}{d\tau} - p_i \dot{x}_0^j, \qquad F_{0i} = \frac{dp_{0i}}{d\tau} - p_j \dot{g}_i^{\ j}. \qquad (5.69)$$

Hence, we see that the equations that we derive from the vanishing of the Spencer derivative of the momenta represent the usual type of mechanical principles in an inertial frame, up to a sign in the generalized torques and forces.

We are now in a position to describe the sense in which our group of motions defines a group of symmetries of a system of differential equations. If we consider the fact that $D^*\phi = 0$ iff $D^*\phi_G = 0$ and $D^*\phi_U = 0$ then we see that the equation in $\phi_G$ defines a class of transformations $g: [a, b] \to G$ by the constraint that $(j^{k-1}g)^*\phi_G = 0$. These transformations then take an integrable initial dynamical state $\phi_U$ to another integrable dynamical state $\phi$. Similarly, if $\psi_0: [a, b] \to J^k([a, b], U)$ is an initial solution to the equation $D^*\phi_U = 0$ – that is, $\psi_0^*(D^*\phi_U) = 0$ – then the transformations $j^{k-1}g$ that satisfy the integrability constraint will take $\psi_0$ to kinematical states $\psi$ that satisfy $\psi^*(D^*\phi) = 0$. Hence, the transformations in question are indeed symmetries of the differential equation that governs the dynamical states.

If we wish to examine the form that our dynamical principle takes in a non-inertial frame, we mostly have to convert $\dot{g}$ into $g\omega$ and use $\theta$ in place of $dg$. This makes:

$$\begin{aligned}
\phi_G|_{D\psi} &= \tilde{T}_j^{\ i} D\theta_i^j + \mathcal{L}_j^i D\omega_i^j \\
&= -\frac{d}{d\tau}\phi_G\Big|_\psi - D^*(\tilde{T}_i^{\ j}\theta_j^i) - D^*(\mathcal{L}_i^j d\omega_j^i).
\end{aligned} \qquad (5.70)$$

The 1-form $\phi_G|_{D\psi}$ remains unchanged in form, although the components $\tilde{F}_{0i}$ can be expressed in terms of the $\omega$'s now.

Hence, to first order, the integrability conditions on the dynamical state $\phi$ in this case take form:

$$\tilde{T}_i^{\ j} = \frac{d\mathcal{L}_i^j}{d\tau}, \qquad \tilde{F}_{0i} = \frac{dp_{0i}}{d\tau}. \qquad (5.71)$$

From the definitions (5.47a, b) of $\tilde{T}_i^{\ k}$ and $\tilde{F}_{0i}$, we can also express them as:

$$T_i^{\ j} = \frac{d\mathcal{L}_i^j}{d\tau} - \mathcal{L}_i^k \omega_k^j - p_{0i}\dot{x}_0^j, \qquad F_{0i} = \frac{dp_{0i}}{d\tau} - p_{0j}\omega_i^j. \qquad (5.72)$$

We see that the equations for the time evolution of the generalized angular momentum then take the form of Euler's equations for rotational motion, while the equations for linear momentum generalize Newton's second law in a non-inertial frame.

If we introduce the notation:

$$\nabla \mathcal{L}_i^j = \frac{d\mathcal{L}_i^j}{d\tau} - \mathcal{L}_i^k \omega_k^j \,, \qquad\qquad \nabla p_{0i} = \frac{dp_{0i}}{d\tau} - p_{0j} \omega_i^j \tag{5.73}$$

then we can put (5.72) into the form:

$$\mathcal{T}_i^{\ j} = \nabla \mathcal{L}_i^j - p_{0i} \dot{x}_0^j \,, \qquad\qquad F_{0i} = \nabla p_{0i} \,. \tag{5.74}$$

Hence, we see that there is an operator that acts on dynamical elements in $V^*(J^{k-1}([a,\,b]$, $\mathfrak{g}[k-1]))$ that is dual to the generalized covariant derivative operator that we introduced on kinematical elements in $J^k([a,\,b]$, $\mathfrak{g}[k])$ in (5.4).

### 5.5  The role of exactness in dynamical states

We now return to the issue of exactness for the 1-form part of our dynamical state $\phi$.

Exactness implies closedness, so if $\phi$ is exact then the components of $\phi$ must satisfy certain conditions that are based in the vanishing of $d\phi$, namely:

$$0 = d\phi = dF_i \wedge dx^i + dp_i \wedge d\dot{x}^i + \cdots + \overset{(k-1)}{dp_i} \wedge \overset{(k-1)}{dx^i} \,. \tag{5.75}$$

One must keep in mind that generally the components of $\phi$ are functions of all the components of the kinematical state $\psi$, so it is not necessary that each term of (5.75) vanish, but only sufficient.

In the case of the first term, $F = F_i \, dx^i$, when $F_i = F_i(x^j, \ \dot{x}^j)$ if $F$ is closed then one has:

$$0 = dF = -\tfrac{1}{2}(F_{i,j} - F_{j,i}) \, dx^i \wedge dx^j - \tfrac{1}{2}\left(\frac{\partial F_i}{\partial \dot{x}^j} - \frac{\partial F_j}{\partial \dot{x}^i}\right) dx^i \wedge d\dot{x}^j \,, \tag{5.76}$$

which gives the local system of partial differential equations for the integrability of the $F_i$ in this sense:

$$0 = \frac{\partial F_i}{\partial x^j} - \frac{\partial F_j}{\partial x^i} = \frac{\partial F_i}{\partial \dot{x}^j} - \frac{\partial F_j}{\partial \dot{x}^i} \,. \tag{5.77}$$

Unless $M$ is simply connected, this necessary condition is not sufficient, though. In a non-simply connected configuration space it is possible for the work done by $F$ around a loop to be non-zero, even though $dF = 0$. As an example, consider the work done by a time-varying magnetic field on a charged particle in a loop that is linked by the magnetic field. In effect, the presence of the magnetic field makes the loop non-contractible, as in the Bohm-Aharonov experiment.

This type of integrability – viz., the integrability of a 1-form relative to exterior differentiation – takes another form in mechanics, namely, the question of whether the differential increments of energy are path-independent for a given choice of endpoints.

This is equivalent to saying that their integral vanishes for any loop (i.e., 1-cycle) $z_1$. This implies that the increment in question must be an exact 1-form. For instance, one might have:

$$F_i \, dx^i = -dU, \qquad p_i \, d\dot{x}^i = d(KE) . \qquad (5.78)$$

The second condition is generally satisfied in point mechanics, since one usually sets:

$$p_i = m \, \delta_{ij} \dot{x}^i , \qquad KE = \tfrac{1}{2} m \, \delta_{ij} \dot{x}^i \dot{x}^j . \qquad (5.79)$$

However, the first condition in (5.78) depends upon the nature of the force. In particular, exactness of the 1-form $F = F_i \, dx^i$ is equivalent to not only the path-independence of the work integral, but to the conservation of energy, which follows from the fact the since $p_i \, d\dot{x}^i$ is usually assumed to be exact, in order for the total 1-form $\phi = F_i \, dx^i + p_i \, d\dot{x}^i$ to be exact – so $\phi = d(KE - U)$ – one must have the exactness of $F$. If $F$ is exact then one calls the force $F_i$ *conservative*.

Not all forces are conservative, though. Two elementary counter-examples are Coulomb friction and viscous drag. In the former case, the magnitude of the force is a constant and the direction is minus direction of the velocity, and in the latter the force itself is minus a constant times the velocity:

$$F_i = -b\dot{x}^i . \qquad (5.80)$$

As a result, one has that $F_i$ satisfies (5.76), so it is closed as a 1-form on $J^1([a, b], M)$, but not exact, unless one has a "velocity potential," which is usually more of an issue in continuum mechanics than in point mechanics. One must also note that there is a clear difference in this case between being exact as a 1-form on $J^1([a, b], M)$ and exact as a 1-form on $M$, since one usually regards forces as 1-forms on the configuration manifold in mechanics, not on the kinematical space.

Of course, the same considerations that apply to generalized forces and momenta also apply to the generalized torques and angular momenta. In that case, one often sets:

$$\mathcal{L}_i^j = I_{ik}^{jl} \omega_l^k , \qquad KE = \tfrac{1}{2} I_{ik}^{jl} \omega_j^i \omega_l^k , \qquad (5.81)$$

in order to make:

$$\mathcal{L}_i^j \, d\omega_j^i = d(KE), \qquad (5.82)$$

If the generalized torque is conservative then one will have a generalized torque potential $U_\tau$ that makes the work done by generalized torque exact:

$$\mathcal{T}_i^{\ j} \theta_j^i = dU_\tau . \qquad (5.83)$$

Another dynamical 1-form whose exactness is often of issue in classical mechanics the 1-form $\mathcal{L}\,d\tau$ itself. By definition, in order for $\mathcal{L}\,d\tau$ to be exact there would have to be an *action function* $S: J^k([a, b], M) \to \mathbb{R}$ such that:

$$\mathcal{L}\,d\tau = dS. \tag{5.84}$$

By Stokes's theorem, the action functional itself would have to be path-independent and one could set:

$$S[a, \tau] = S(\tau) - S(a). \tag{5.85}$$

Generally, one first converts $\mathcal{L}\,d\tau$ into the Poincaré-Cartan form $\Theta = p_i\,dx^i - H(\tau, x^i, p_i)\,d\tau$ by means of the Legendre transform and then examines the consequences of assuming that $\Theta$ is exact, that is $\Theta = dS$. By differentiation, we obtain the Hamilton-Jacobi equations for $S$ as a necessary and sufficient condition for the exactness of $\Theta$:

$$p_i = \frac{\partial S}{\partial x^i}, \qquad H(\tau, x^i, \frac{\partial S}{\partial x^i}) = -\frac{\partial S}{\partial \tau}. \tag{5.86}$$

## 5.6 Constitutive laws

At first, our way of associating dynamical states with kinematical ones was by defining a Lagrangian function $\mathcal{L}$ on the kinematical state space and calling the vertical part of $\phi = d\mathcal{L}|_\psi$ the dynamical state that is associated with the kinematical state $\psi$. We then pointed out that there are more general vertical 1-forms on $J^{k-1}([a, b], M)$ or $J^{k-1}([a, b], G) \times J^{k-1}([a, b], U)$ that one could use to represent dynamical states with than the just the exact ones.

In practice, the process of associating dual dynamical objects to kinematical objects often involves the introduction of a *mechanical constitutive law*. Actually, we are going to enlarge the scope of that term to include not only the association of forces with displacements, but also the association of momenta with velocities.

If one writes out the components of a general $\phi \in T[a, b] \otimes V(J^{k-1}([a, b], M)$ as functions of the coordinates of $J^{k-1}([a, b]; M)$:

$$F_i = F_i(\tau, x^i, \dot{x}^i, \ldots, \overset{(k-1)}{x^i}), \qquad p_i = p_i(\tau, x^i, \dot{x}^i, \ldots, \overset{(k-1)}{x^i}), \qquad \ldots \tag{5.87}$$

then one sees that what these functions represent are generalized constitutive laws.

Thus, we can think of a general (nonlinear) constitutive law as a smooth section $\chi$: $J^{k-1}([a, b], M) \to V^*(J^{k-1}([a, b], M))$ that takes its values in the 1-forms that are vertical for the projection of $J^{k-1}([a, b], M)$ onto $[a, b]$; then again, this is also how we defined a dynamical state. The way that one distinguishes one from the other is that, in practice, a constitutive law is defined by a specific set of functional relationships of the form (5.87), so it is really just the difference between choosing an arbitrary section and choosing a specific one.

Commonly, one deals with simpler forms of these functions than (5.87) suggests. For instance, in a second-order mechanical model, for which there are no further kinematical or dynamical components than the ones mentioned specifically in (5.87), it is common to use laws of the form:

$$F_i = F_i(x^i), \qquad p_i = p_i(\dot{x}^i), \tag{5.88}$$

such as when one has a time-independent force acting on a time-invariant mass.

One might include velocity-dependent dissipative forces and time-varying mass, as in the problem of jet propulsion in an aerodynamic medium: The functions then take the form:

$$F_i = F_i(x^i, \dot{x}^i), \qquad\qquad p_i = p_i(\tau, \dot{x}^i). \tag{5.89}$$

When we do the same thing for the components of $\phi \in \Lambda^1(T^{k-1}(G))$:

$$\mathcal{T}_i^{\ j} = \mathcal{T}_i^{\ j}(\tau, g, \omega, \cdots, \overset{(k-2)}{\omega}), \qquad\qquad L_i^j = L_i^j(\tau, g, \omega, \cdots, \overset{(k-2)}{\omega}), \tag{5.90}$$

we see that the most illuminating way of describing such a law is that it involves a one-to-one correspondence $C: \mathfrak{g} \to \mathfrak{g}^*$ between the elements of $\mathfrak{g}$ and the elements of $\mathfrak{g}^*$.

The general term for an element of $\mathfrak{g}^*$ is *torsor* [**1, 5, 6**]. One sees that they include both forces and linear momenta when $\mathfrak{g} = \mathbb{R}^n$, as well as torques and angular momenta when $\mathfrak{g} = \mathfrak{so}(n)$. In any event, if the elements of $\mathfrak{g}$ are regarded as the infinitesimal generators of motions then the evaluation of a torsor on an infinitesimal motion gives a differential increment of energy in one form or another.

We again point out that the constructions above are not specific to the rotation group, but apply just as well to Lorentz transformations and the case of the "pseudo-rigid" body.

Often a constitutive law is assumed to be linear, but it is important to understand that the most common origin of nonlinearity in physics is the breakdown of linearity in a constitutive law when the magnitude of the kinematical object – say, the displacement vector field – exceeds some practical limit. For instance, Hooke's law $F = -kx$ is simply an empirical approximation that only applies to elastic materials when the displacement is small. One can also observe that inverse-square laws of force essentially define nonlinear associations of forces with translations. However, as illustrated by (5.79) and (5.81), linear constitutive laws for associating momenta with velocities are commonplace in mechanics.

The question then arises whether the introduction of a constitutive law is more or less general than the introduction of a Lagrangian. Here, we remind the reader of the previous discussion of the limits of Lagrangian methodology and the fact that there are more general 1-forms on $J^{k-1}([a, b]; U)$ or $J^{k-1}([a, b]; G)$ than the exact 1-forms that are obtained from the differentials of Lagrangians. In such a case – for instance, viscous damping − one usually resorts to the use of a constitutive law, anyway. Also, many Lagrangians are constructed by starting with a constitutive law for force and the usual one for momentum and using $T(\dot{x}) - U(x)$ for a Lagrangian, where $T$ refers to the kinetic energy and $U$ to the potential energy function. Hence, there is good reason to consider the definition of a constitutive law, together with integrability conditions on the dynamical states, as a more general way of formulating equations of motion for a mechanical model than the Lagrangian formulation.

## 6 Examples

Let us illustrate these concepts in the case of $U \subset \mathbb{R}^n$ for three common group actions: the action of $\mathbb{R}^n$ on $U$ by translations, the linear action of the rotation group $SO(n)$ on $U$, and the linear action of the Lorentz group on $U$ when $\mathbb{R}^n$ is four-dimensional Minkowski space. We shall also use the proper time interval $[0, 1]$, for specificity.

### 6.1 Translational motion

The action of $\mathbb{R}^n$ on $U$ by translations is simply $\mathbb{R}^n \times U \to \mathbb{R}^n$, $(s^i, x_0^i) \mapsto x^i$, where:

$$x^i = x_0^i + s^i. \tag{6.1}$$

The prolongation of this action to $J^k([0, 1], \mathbb{R}^n) \times J^k([0, 1], U) \to J^k([0, 1], \mathbb{R}^n)$ is obtained by differentiation of (6.1) (while treating $x_0^i$ as a function of $\tau$) :

$$\overset{(k)}{x^i}(\tau) = \overset{(k)}{x_0^i} + \overset{(k)}{s^i}(\tau) \tag{6.2}$$

One has to be somewhat careful in interpreting this equation since the action of $J^k([0, 1], \mathbb{R}^n)$ by translation of the higher-order derivatives is only infinitesimal and $\overset{(k)}{x^i}$ differs from $\overset{(k)}{x_0^i}$ by a finite time interval. However, for an integrable section of $J^k([0, 1], \mathbb{R}^n)$ one has:

$$d \overset{(k)}{x^i} = \overset{(k+1)}{s^i} d\tau . \tag{6.3}$$

The finite form of the translation is then obtained by integration:

$$\overset{(k)}{x^i}(\tau) = \overset{(k)}{x_0^i} + \int_0^\tau \overset{(k+1)}{s^i}(\sigma) d\sigma = \overset{(k)}{x_0^i} + \overset{(k)}{s^i}(\tau) , \tag{6.4}$$

which is consistent with (6.2).

Hence, we regard a kinematical state, in one sense, as a section $s^{(k)}$: $[0, 1] \to J^k([0, 1],$ $\mathbb{R}^n$), $\tau \mapsto (\tau, s^i(\tau), \ldots \overset{(k)}{s^i}(\tau))$ that acts on an element $\psi_0 \in (0, x_0^i, \ldots, \overset{(k)}{x_0^i})$ in $J^k([0, 1], U)$, to produce a kinematical state, in the sense of a section $\psi$: $[0, 1] \to J^k([0, 1], \mathbb{R}^n)$, $\tau \mapsto (\tau,$ $x^i(\tau), \ldots \overset{(k)}{x^i}(\tau))$.

A kinematical state $\psi(\tau) = (\tau, x^i(\tau), \overset{(1)}{x^i}(\tau), \ldots, \overset{(k)}{x^i}(\tau))$ is integrable iff:

$$0 = D\psi = \left[\left(\frac{dx^i}{d\tau} - \overset{(1)}{x^i}\right), \cdots, \left(\frac{d\,\overset{(k-1)}{x^i}}{d\tau} - \overset{(k)}{x^i}\right)\right] d\tau, \qquad (6.5)$$

which simply states that each successive set of components $\overset{(l)}{x^i}(\tau)$ is the proper time derivative of the previous one for $l = 1, \ldots, k - 1$.

Similarly, the integrability of the kinematical state $s^{(k)}(\tau) = (\tau, s^i(\tau), \ldots \overset{(k)}{s^i}(\tau))$ is equivalent to:

$$0 = D\,s^{(k)} = \left[\left(\frac{ds^i}{d\tau} - \overset{(1)}{s^i}\right), \cdots, \left(\frac{d\,\overset{(k-1)}{s^i}}{d\tau} - \overset{(k)}{s^i}\right)\right] d\tau, \qquad (6.6)$$

which makes an analogous statement to the one implied by (6.5)

From (6.3), $d\,\overset{(k)}{x^i}$ pulls back to $dx_0^{(k)} + ds^{(k)}$ under the action of translation. As a result, if a dynamical state is represented by a vertical 1-form $\phi$ on $T([a, b]) \otimes J^{k-1}([0, 1], \mathbb{R}^n)$, i.e., a 1-form on $T^{k-1}(\mathbb{R}^n)$ of the form $F_i\,dx^i + p_i\,d\dot{x}^i + \ldots + \overset{(k-2)}{p_i}\,d\overset{(k-1)}{x^i}$ then $\phi$ pulls back to $\phi_s + \phi_0$, with:

$$\phi_s = \left(F_i ds^i + p_i ds^i + \cdots \overset{(k-2)}{p_i}\,\overset{(k-1)}{ds^i}\right) \otimes \frac{\partial}{\partial \tau}, \qquad (6.7b)$$

$$\phi_0 = \left(F_{0i}dx_0^i + p_{0i}d\dot{x}_0^i + \cdots + \overset{(k-2)}{p_{0i}}\,\overset{(k-1)}{dx_0^i}\right) \otimes \frac{\partial}{\partial \tau}. \qquad (6.7b)$$

The integrability of the dynamical state $\phi$ is equivalent to:

$$0 = D^*\phi = \left(F_i - \frac{dp_i}{d\tau}\right)dx^i + \left(p_i - \frac{d\,\overset{(1)}{p_i}}{d\tau}\right)d\dot{x}^i + \cdots + \left(\overset{(k-3)}{p_i} - \frac{d\,\overset{(k-2)}{p_i}}{d\tau}\right)d\,\overset{(k-2)}{x^i}, \qquad (6.8)$$

which gives Newton's equations when $k = 2$.

Similarly, the integrability of the dynamical state $\phi_s$ is equivalent to:

$$0 = D^* \phi_s = \left( F_i - \frac{dp_i}{d\tau} \right) ds^i + \left( p_i - \frac{d \overset{(1)}{p_i}}{d\tau} \right) d\dot{s}^i + \cdots + \left( \overset{(k-3)}{p_i} - \frac{d \overset{(k-2)}{p_i}}{d\tau} \right) \overset{(k-2)}{d s^i} , \qquad (6.9)$$

which implies the same conclusion.

One sees that, in effect, the role of the initial state is entirely passive in the case of translational motion. This is nothing but the observation that the initial state simply represents a set of integration constants for the motion, and the derivative of any constant is zero.

### 6.2  Rotational motion

Now, suppose that $G = SO(n)$, where $n = 2$ or $3$ in non-relativistic rotational mechanics. The main difference between the two cases, as observed above, is that for $n = 2$ one is dealing with an Abelian Lie group and for $n = 3$ the Lie group is non-Abelian.

Some adjustments to the general notation can be made to account for the fact that the elements of $\mathfrak{so}(n)$ are all anti-symmetric matrices. First, one sets:

$$dg^i_j = g^i_k \omega^k_j d\tau = g^i_k \theta^k_j , \qquad (6.10)$$

in which the matrix $\theta^i_j$ of Maurer-Cartan 1-forms will be anti-symmetric. Indeed, in the case of $n = 2$, it will take the form of $J^i_j d\theta$, where $J^i_j$ is the elementary anti-symmetric 2×2 matrix. The $d\theta$ will represent a differential increment of angle in a plane that is perpendicular to the axis of rotation, although again we point out that the 1-form $d\theta$, despite the popularity of the notation, is not actually exact.

Let us re-examine equations (5.7) for the velocity and acceleration as they are described in terms of the present group action:

$$\dot{x}^i = g^i_j (\omega^j_k x^k_0 + \dot{x}^j_0) = g^i_j \nabla x^j_0 , \qquad (6.11a)$$

$$\ddot{x}^i = g^i_j (\overset{(1)}{\omega^j_k} x^k_0 + 2\omega^j_k \dot{x}^k_0 + \ddot{x}^j_0) = g^i_j \nabla^2 x^j_0 . \qquad (6.11b)$$

in which:

$$\overset{(1)}{\omega^i_j} = \dot{\omega}^i_j + \omega^i_k \omega^k_j . \qquad (6.12)$$

One sees in these expressions the usual Coriolis contributions, $\omega^j_j x^j_0$ and $2\omega^j_j \dot{x}^j_0$, resp., to the velocity and acceleration, resp., along with the normal acceleration $\dot{\omega}^j_j x^j_0$ and centripetal acceleration $\omega_k^i \omega_j^k x_0^j$. The $\nabla$ operator clearly asserts itself as what often gets called the "rotational derivative" operator in rotational mechanics.

From (5.36a, b), we have, to second order ($k = 2$):

$$\phi_G = \tilde{T}_k^{\ j} \theta_j^i + \mathcal{L}_i^j d\omega_j^i \tag{6.13}$$

and:

$$\phi_U = (F_{0i} + p_{0j}\omega_i^j)dx_0^i + p_{0i}d\dot{x}_0^i, \tag{6.14}$$

in which, from (5.34) and (5.37):

$$\tilde{T}_i^{\ j} = T_i^{\ j} + \mathcal{L}_i^k \omega_k^j + p_{0i}\dot{x}_0^j, \qquad T_i^{\ j} = F_{0i}x_0^j, \qquad \mathcal{L}_i^j = p_{0i}x_0^j. \tag{6.15}$$

When we sum over all $i$ and $j$ in the torque work 1-form $\tilde{T} = \tilde{T}_k^{\ j}\theta_j^i$, as well as the rotational kinetic energy 1-form $\mathcal{L} = \mathcal{L}_i^j d\omega_j^i$, only the anti-symmetric parts of $\tilde{T}_i^{\ j}$ and $\mathcal{L}_i^j$ will contribute to the sum:

$$\tilde{T} = \tfrac{1}{2}(\tilde{T}_i^{\ j} - \tilde{T}_j^{\ i})\theta_j^i, \qquad\qquad \mathcal{L} = \tfrac{1}{2}(\mathcal{L}_i^j - \mathcal{L}_j^i)d\omega_j^i. \tag{6.16}$$

If we explicitly expand the components in the right-hand sides of these expressions then we get:

$$\tilde{T}_i^{\ j} = \tfrac{1}{2}(F_{0i}x_0^j - F_{0j}x_0^i) + \tfrac{1}{2}(\mathcal{L}_i^k\omega_k^j - \mathcal{L}_j^k\omega_k^i) + \tfrac{1}{2}(p_{0i}\dot{x}_0^j - p_{0j}\dot{x}_0^i), \tag{6.17a}$$

$$\mathcal{L}_i^j = \tfrac{1}{2}(p_{0i}x_0^j - p_{0j}x_0^i). \tag{6.17b}$$

Since we are assuming that both $\tilde{T}_i^{\ j}$ and $\mathcal{L}_i^j$ are elements of $\mathfrak{so}(n)^*$, they will be anti-symmetric to begin with so there is no abuse of notation associated with using them for the left-hand sides of (6.17a, b). Furthermore, from the anti-symmetry of $\omega_j^i$ we can also express the second term in the right-hand side of (6.17a) as:

$$\tfrac{1}{2}(\mathcal{L}_i^k - \mathcal{L}_k^i)\omega_k^j = \mathcal{L}_i^k\omega_k^j. \tag{6.18}$$

The expressions on the right-hand sides of (6.17a, b) are more similar to the forms that one obtains in rotational mechanics by the use of cross products, up to sign. Of course, that is because the cross product endows $\mathbb{R}^3$ with the structure of a Lie algebra that is isomorphic to $\mathfrak{so}(3)$ by the adjoint map that takes any $\mathbf{v} \in \mathbb{R}^3$ to the anti-symmetric matrix:

$$\text{ad}(\mathbf{v}) = v^i J_i = \begin{bmatrix} 0 & -v^3 & v^2 \\ v^3 & 0 & -v^1 \\ -v^2 & v^1 & 0 \end{bmatrix}, \tag{6.19}$$

in which $J_i$, $i = 1, 2, 3$, are the elementary anti-symmetric 3×3 matrices, which are the infinitesimal generators of the finite rotations that are described by the Euler angles.

One notes that the contribution from the last term in (6.17a) will vanish when $p_{0i} = m\delta_{ij}\dot{x}_0^j$. Hence, it is only relevant to the case of "transversal momenta," i.e., momenta that are not collinear with covelocity.

From (5.56b) and (5.60), the integrability of $\phi$ is equivalent to:

$$\mathcal{T}_i^{\ j} = \frac{d\mathcal{L}_i^j}{d\tau} - \mathcal{L}_i^k \omega_k^j - \tfrac{1}{2}(p_{0i}\dot{x}_0^j - p_{0i}\dot{x}_0^j), \qquad F_{0i} = \frac{dp_{0i}}{d\tau} - p_{0j}\omega_i^k. \tag{6.20}$$

The first of these equations represents Euler's equations for the time evolution of angular momentum when viewed in a rotating frame, along with a contribution from transverse momentum, if there is one. The second equation is the form that Newton's second law takes in a rotating frame.

We can also use the $\nabla$ operator to put these equations in the form:

$$\mathcal{T}_i^{\ j} = \nabla \mathcal{L}_i^j - \tfrac{1}{2}(p_{0i}\dot{x}_0^j - p_{0i}\dot{x}_0^j), \qquad F_{0i} = \nabla p_{0j}. \tag{6.21}$$

This, too, is consistent with the interpretation of $\nabla$ as the rotational derivative operator.

### 6.3 Lorentz transformations

In order to make things more relativistic, one mostly needs to do two things: First, one must recognize that although the Lie algebra $\mathfrak{so}(3, 1)$ does not consist of anti-symmetric 4×4 matrices, nevertheless, since any $\omega_j^i \in \mathfrak{so}(3, 1)$ must satisfy $\omega = -\omega^*$, which is equivalent to the component form:

$$\eta^{ik}\omega_k^j = -\eta^{jk}\omega_k^i, \tag{6.22}$$

we can say that the matrix $\omega^{ji} = \eta^{ik}\omega_k^j$ that is associated with $\omega_j^i$ is anti-symmetric.

One can then rearrange the indices in (6.13) accordingly to obtain:

$$\tilde{\mathcal{T}} = \tilde{\mathcal{T}}_{ij}\theta^{ij}, \qquad \mathcal{L} = \mathcal{L}_{ij}\,d\omega^{ij}, \tag{6.23}$$

in which:

$$\tilde{\mathcal{T}}_{ij} = \tfrac{1}{2}(F_{0i}x_{0j} - F_{0j}x_{0i}) + \mathcal{L}_{ik}\omega_j^k + \tfrac{1}{2}(p_{0i}\dot{x}_{0j} - p_{0j}\dot{x}_{0i}), \tag{6.24a}$$

$$\mathcal{L}_{ij} = \tfrac{1}{2}(p_{0i}x_{0j} - p_{0j}x_{0i}). \tag{6.24b}$$

In this form, the expressions in parentheses take the form of the components of the exterior products $F_0 \wedge x_0$, $p_0 \wedge \dot{x}_0$, and $p_0 \wedge x_0$ of the 1-forms $F_0 = F_{0i}\,dx_0^i$, $x_0 = x_{0i}\,dx_0^j$, $\dot{x}_0 = \dot{x}_{0i}dx_0^i$, $p_0 = p_{0i}\,dx_0^i$.

The integrability equations (5.69) for $\phi$ can then be put into the form:

$$\mathcal{T}_{ij} = \frac{d\mathcal{L}_{ij}}{d\tau} - \mathcal{L}_{ik}\omega_j^k - (p_0 \wedge \dot{x}_0)_{ij}, \qquad F_{0i} = \frac{dp_{0i}}{d\tau} - p_{0j}\omega_i^j. \qquad (6.25)$$

The second relativistic consideration is that one must account for the fact that not all possible velocity vectors are physically meaningful, but only the ones that lie on the unit proper time hyperboloid:

$$1 = \eta(\dot{x}, \dot{x}) = \eta_{ij}\dot{x}^i\dot{x}^j, \qquad (6.26)$$

which is true for the tangent vectors to $M = \mathbb{R}^4$ whether they are used in the form $\dot{x}(\tau)$ or in the form $\dot{x}_0^i$.

A sequence of consequences follow from (6.26) by differentiation:

$$0 = \eta(\dot{x}, \ddot{x}) = \eta(\ddot{x}, \ddot{x}) + \eta(\dot{x}, \dddot{x}) = \dots \qquad (6.27)$$

which one can express as:

$$0 = \eta_{ij}\dot{x}^i\ddot{x}^j = \eta_{ij}(\ddot{x}^i\ddot{x}^j + \dot{x}^i\dddot{x}^j) = \dots \qquad (6.28)$$

These conditions have the effect of further restricting the $k$-jets that define physically acceptable kinematical states beyond their integrability to require them to lie on a quadratic hypersurface in $J^k([a, b], M)$.

In terms of the action of the Lorentz group on tangent vectors, one sees that if $\dot{x} = g(\omega x_0 + \dot{x}_0)$ is restricted to the unit hyperboloid, along with $\dot{x}_0$, then if $g$ takes its values in the Lorentz group one can say that $\omega x_0 + \dot{x}_0$ also lies on the unit hyperboloid, which gives:

$$1 = \eta(\omega x_0, \omega x_0) + \eta(\omega x_0, \dot{x}_0) + \eta(\dot{x}_0, \dot{x}_0), \qquad (6.29)$$

which then reduces to:

$$0 = \eta(\omega x_0, \omega x_0) + \eta(\omega x_0, \dot{x}_0^i). \qquad (6.30)$$

Hence, for a given initial kinematical state $(x_0, \dot{x}_0)$ the only causal $\omega$'s will lie on a quadratic hypersurface in $\mathfrak{so}(3, 1)$.

One can also find a hypersurface for acceptable $\overset{(1)}{\omega}$'s for a given initial kinematical state $(x_0, \dot{x}_0, \ddot{x}_0)$, and so forth by a similar process, although the details become rapidly tedious.

Dually, a causal energy-momentum 1-form $p$ must lie on the "mass shell":

$$m_0^2 c^2 = \eta(p, p) = \eta^{ij} p_i p_j,\qquad\qquad (6.31)$$

in which $m_0$ represents the rest mass of the particle in question.

By differentiation, this gives a causality constraint on force 1-forms:

$$0 = \eta(p, F) = \eta^{ij} p_i F_j.\qquad\qquad (6.32)$$

Note that the physical significance of further differentiations is lost in the space $V^*(J^{k-1}([a, b], M))$ since proper time differentiation reduces the order of the components of a vertical 1-form on $J^{k-1}([a, b], M)$ instead of increasing it.

## 7 Summary

To summarize the basic points of the foregoing, we state:

1. When the kinematical state of a moving point in a configuration manifold $M$ is represented by a section $\psi(\tau)$ of the bundle $J^k([a, b], M) \to [a, b]$, the most elementary form of physical motion is described by an integrable section.. This condition can be expressed concisely in terms of the Spencer operator as:

$$D\psi = 0.$$

However, non-integrable motions are still physically realizable.

2. If the motion is a result of the action of a Lie group $G$ of transformations on $U \subset M$ then the kinematical state $\psi$ can also be represented by the pair $(\psi_G, \psi_U)$, where $\psi_G$ is a section of the bundle $J^k([a, b], G) \to [a, b]$ and $\psi_U$ is an element of the fiber of $J^k([a, b], U) \to [a, b]$ over $a$ that represents an initial kinematical state. However, the integrability of $\psi$ is not equivalent to the vanishing of both $D\psi_G$ and $D\psi_U$, since the latter condition is true only for the initial kinematical state that represents a state of rest. Hence, one must consider an integrability condition that takes the form:

$$0 = D\psi_G \cdot \psi_G + \psi_G \cdot D\psi_G$$

in order to make $\psi = \psi_G \cdot \psi_U$ integrable. Conversely, if $\psi_G$ is an integrable section of $J^k([a, b], G) \to [a, b]$ and $\psi_U$ is an arbitrary initial kinematical state then the resulting kinematical state $\psi$ will not generally be integrable.

3.   The manifold $J^k([a, b], G)$ can be represented in either the form $[a, b] \times T^k(G)$, which represents a generalization of rotational mechanics relative to an inertial frame, or as $[a, b] \times G \times \mathfrak{g}[k]$, which generalizes rotational mechanics with respect to a co-moving non-inertial frame.

4.   The representation of dynamical states by vertical 1-forms $\phi$ on $T([a, b]) \otimes J^{k-1}([a, b], M)$ is a natural generalization from the customary constructions that one makes in the variational formulation of mechanics, which only accounts for the exact forms. Similarly, the resulting integrability condition that one imposes on $\phi$, namely:

$$D^*\phi = 0,$$

also represents a natural generalization of the Newtonian and Lagrangian formulation of mechanics that is still valid for non-conservative forces.

5.   The corresponding form $\phi_G + \phi_U$ that $\phi$ takes when it has been pulled back along the group action $G \times U \to M$ is a natural generalization of the usual constructions that one makes in rotational mechanics to more general groups.  As opposed to the situation in kinematics, the integrability of $\phi$ is equivalent to the vanishing of both $\phi_G$ and $\phi_U$.  When these conditions:

$$D^*\phi_G = 0, \qquad D^*\phi_U = 0,$$

are expressed in terms of 1-forms on $T^{k-1}(G)$, the resulting equations generalize the form of Newton's second law for both linear and angular momentum in an inertial frame. When they are expressed in terms of 1-forms on $G \times \mathfrak{g}[k-1]$, the resulting local equations represent a natural generalization of the Euler equations for motion in a rotating frame, along with the form that Newton's second law takes in that frame.

Furthermore, the above pair of differential equations allows one to describe the action of $G$ by saying that the first equation defines a class of prolongations of the maps $g$: $[a, b] \to G$ that take solutions of the dynamical equation $D^*\phi = 0$ to other solutions; i.e., they are symmetries of that differential equation.

6.   The association $\phi = \phi(\psi)$ of a dynamical state with a kinematical state is most generally defined by a set of mechanical constitutive laws, which amounts to specifying particular functional forms for the components of the dynamical state.  This process is more general than the process of starting with a Lagrangian function on $J^k([a, b], M)$ which only produces the components of exact 1-forms by differentiation.

Although integrability does not have the intuitive appeal as a statement of natural philosophy that one finds in the least-action principle, nevertheless, because it represents a formal mathematical generalization of the variational methodology, it might suggest a possible generalization of the least-action principle that does have such an intuitive appeal.  Certainly, quantum physics has already given considerable physical evidence that the least-action principle is just another level of approximation in the process of the mathematical modeling of physical phenomena.

In the next Part of this series of articles, we shall go over the constructions and results of sections 4 through 6 for the case of the motion of an extended body $K \subset \mathbb{R}^m$, instead of a pointlike body $[a, b] \subset \mathbb{R}$. We shall see that the methods that we described here are sufficiently robust as to admit a natural extension from $m = 1$ to more general $m$ by essentially replacing total proper time derivatives with partial spacetime derivatives. However, the question of integrability becomes more involved since the Spencer sequence will terminate at a later stage when $m > 1$.